\documentclass[10pt,journal,compsoc]{IEEEtran}
\ifCLASSOPTIONcompsoc
  \usepackage[nocompress]{cite}
\else
  \usepackage{cite}
\fi
\ifCLASSINFOpdf
  \usepackage[pdftex]{graphicx}
\else
\fi
\usepackage{amsmath}
\usepackage{amssymb}
\usepackage{algorithm}
\usepackage{algpseudocode}
\usepackage{array}
\ifCLASSOPTIONcompsoc
  \usepackage[caption=false,font=footnotesize,labelfont=sf,textfont=sf]{subfig}
\else
  \usepackage[caption=false,font=footnotesize]{subfig}
\fi
\def\etal{\emph{et~al}.~}

\begin{document}
%
% paper title
% Titles are generally capitalized except for words such as a, an, and, as,
% at, but, by, for, in, nor, of, on, or, the, to and up, which are usually
% not capitalized unless they are the first or last word of the title.
% Linebreaks \\ can be used within to get better formatting as desired.
% Do not put math or special symbols in the title.
\title{FoR$^2$M: Recognition and Repair of Foldings in Mesh Surfaces. Application to 3D Object Degradation.}
%
%
% author names and IEEE memberships
% note positions of commas and nonbreaking spaces ( ~ ) LaTeX will not break
% a structure at a ~ so this keeps an author's name from being broken across
% two lines.
% use \thanks{} to gain access to the first footnote area
% a separate \thanks must be used for each paragraph as LaTeX2e's \thanks
% was not built to handle multiple paragraphs
%
%
%\IEEEcompsocitemizethanks is a special \thanks that produces the bulleted
% lists the Computer Society journals use for "first footnote" author
% affiliations. Use \IEEEcompsocthanksitem which works much like \item
% for each affiliation group. When not in compsoc mode,
% \IEEEcompsocitemizethanks becomes like \thanks and
% \IEEEcompsocthanksitem becomes a line break with idention. This
% facilitates dual compilation, although admittedly the differences in the
% desired content of \author between the different types of papers makes a
% one-size-fits-all approach a daunting prospect. For instance, compsoc
% journal papers have the author affiliations above the "Manuscript
% received ..."  text while in non-compsoc journals this is reversed. Sigh.

\author{Konstantinos~Sfikas,
        Panagiotis~Perakis,~\IEEEmembership{Member,~IEEE Computer Society,}
        and~Theoharis~Theoharis% <-this % stops a space
%\IEEEcompsocitemizethanks{\IEEEcompsocthanksitem M. Shell was with the Department
%of Electrical and Computer Engineering, Georgia Institute of Technology, Atlanta,
%GA, 30332.\protect\\
% note need leading \protect in front of \\ to get a newline within \thanks as
% \\ is fragile and will error, could use \hfil\break instead.
%E-mail: see http://www.michaelshell.org/contact.html
%\IEEEcompsocthanksitem J. Doe and J. Doe are with Anonymous University.}% <-this % stops an unwanted space

\IEEEcompsocitemizethanks{\IEEEcompsocthanksitem K. Sfikas is with the Hellenic Ministry of Digital Governance,
11 Fragkoudi str., 10163, Athens, Greece. \protect\\
E-mail: k.sfikas@mindigital.gr
\IEEEcompsocthanksitem P. Perakis is an independent researcher
\IEEEcompsocthanksitem T. Theoharis is with the Department of Computer and Information Science,
Norwegian University of Science and Technology (NTNU),
Sem S{\ae}lands Vei 9, 7491, Trondheim, Norway.\protect\\
E-mail: theotheo@ntnu.no}% <-this % stops an unwanted space

%\thanks{Manuscript received April 19, 2005; revised August 26, 2015.}}
}

\IEEEtitleabstractindextext{%
\begin{abstract}

Triangular meshes are the most popular representations of 3D objects, but many mesh surfaces contain topological singularities that represent a challenge for displaying or further processing them properly. One such singularity is the self-intersections that may be present in mesh surfaces that have been created by a scanning procedure or by a deformation transformation, such as off-setting.

Mesh foldings comprise a special case of mesh surface self-intersections, where the faces of the 3D model intersect and become reversed, with respect to the \emph{unfolded} part of the mesh surface. A novel method for the recognition and repair of mesh surface foldings is presented, which exploits the structural characteristics of the foldings in order to efficiently detect the \emph{folded} regions. Following detection, the foldings are removed and any gaps so created are filled based on the geometry of the 3D model. The proposed method is directly applicable to simple mesh surface representations while it does not perform any embedding of the 3D mesh (i.e. voxelization, projection). Target of the proposed method is to facilitate mesh degradation procedures in a fashion that retains the original structure, given the operator, in the most efficient manner.

\end{abstract}

% Note that keywords are not normally used for peerreview papers.
}

% make the title area
\maketitle

% To allow for easy dual compilation without having to reenter the
% abstract/keywords data, the \IEEEtitleabstractindextext text will
% not be used in maketitle, but will appear (i.e., to be "transported")
% here as \IEEEdisplaynontitleabstractindextext when the compsoc
% or transmag modes are not selected <OR> if conference mode is selected
% - because all conference papers position the abstract like regular
% papers do.
\IEEEdisplaynontitleabstractindextext
% \IEEEdisplaynontitleabstractindextext has no effect when using
% compsoc or transmag under a non-conference mode.

% For peer review papers, you can put extra information on the cover
% page as needed:
% \ifCLASSOPTIONpeerreview
% \begin{center} \bfseries EDICS Category: 3-BBND \end{center}
% \fi
%
% For peerreview papers, this IEEEtran command inserts a page break and
% creates the second title. It will be ignored for other modes.
\IEEEpeerreviewmaketitle

\IEEEraisesectionheading{\section{Introduction}\label{sec:introduction}}
% Computer Society journal (but not conference!) papers do something unusual
% with the very first section heading (almost always called "Introduction").
% They place it ABOVE the main text! IEEEtran.cls does not automatically do
% this for you, but you can achieve this effect with the provided
% \IEEEraisesectionheading{} command. Note the need to keep any \label that
% is to refer to the section immediately after \section in the above as
% \IEEEraisesectionheading puts \section within a raised box.

% The very first letter is a 2 line initial drop letter followed
% by the rest of the first word in caps (small caps for compsoc).
%
% form to use if the first word consists of a single letter:
% \IEEEPARstart{A}{demo} file is ....
%
% form to use if you need the single drop letter followed by
% normal text (unknown if ever used by the IEEE):
% \IEEEPARstart{A}{}demo file is ....
%
% Some journals put the first two words in caps:
% \IEEEPARstart{T}{his demo} file is ....

\IEEEPARstart{T}{he} existence of various topological singularities on the mesh surface of 3D models is a frequent phenomenon. Such singularities represent a challenge for displaying or further processing the 3D models properly.

The causes of such defects vary significantly, from data acquisition to data processing procedures, e.g. overlapped range scans, range scan fusion, manually created CAD models, joined NURBS patches, tessellation of CAD models, contoured meshes and deformed meshes.

One type of topological singularity is the self-intersections which mesh surfaces can exhibit when created by a digitization process or by a deformation transformation, such as an off-setting operator. Mesh surface folding is a special case of self-intersections and will be the focus of this paper.

In order to achieve consistency between the structure of the original object and that of the 3D model, surface foldings must: (a) be prevented from occuring on the mesh surface; or (b) be detected and repaired so that the mended mesh surface is as coherent as possible to the original object mesh surface.

The method proposed in this paper focuses on the latter goal: given a mesh surface that exhibits surface foldings, aim at identifying them and subsequently remove the defective areas and reconstruct any gaps that may have occurred.

The proposed method exploits the structural characteristics of the foldings in order to efficiently detect the \emph{folded} regions. Following detection, the foldings are removed and any gaps created are filled based on the geometry of the 3D model. The proposed method is directly applicable to simple mesh surface representations and does not perform any embedding of the 3D mesh (e.g. voxelization, projection).

\begin{figure*}[!ht]
\centering
\begin{minipage}[b]{.66\linewidth}
\centering
    \subfloat{\includegraphics[width=.6\linewidth]{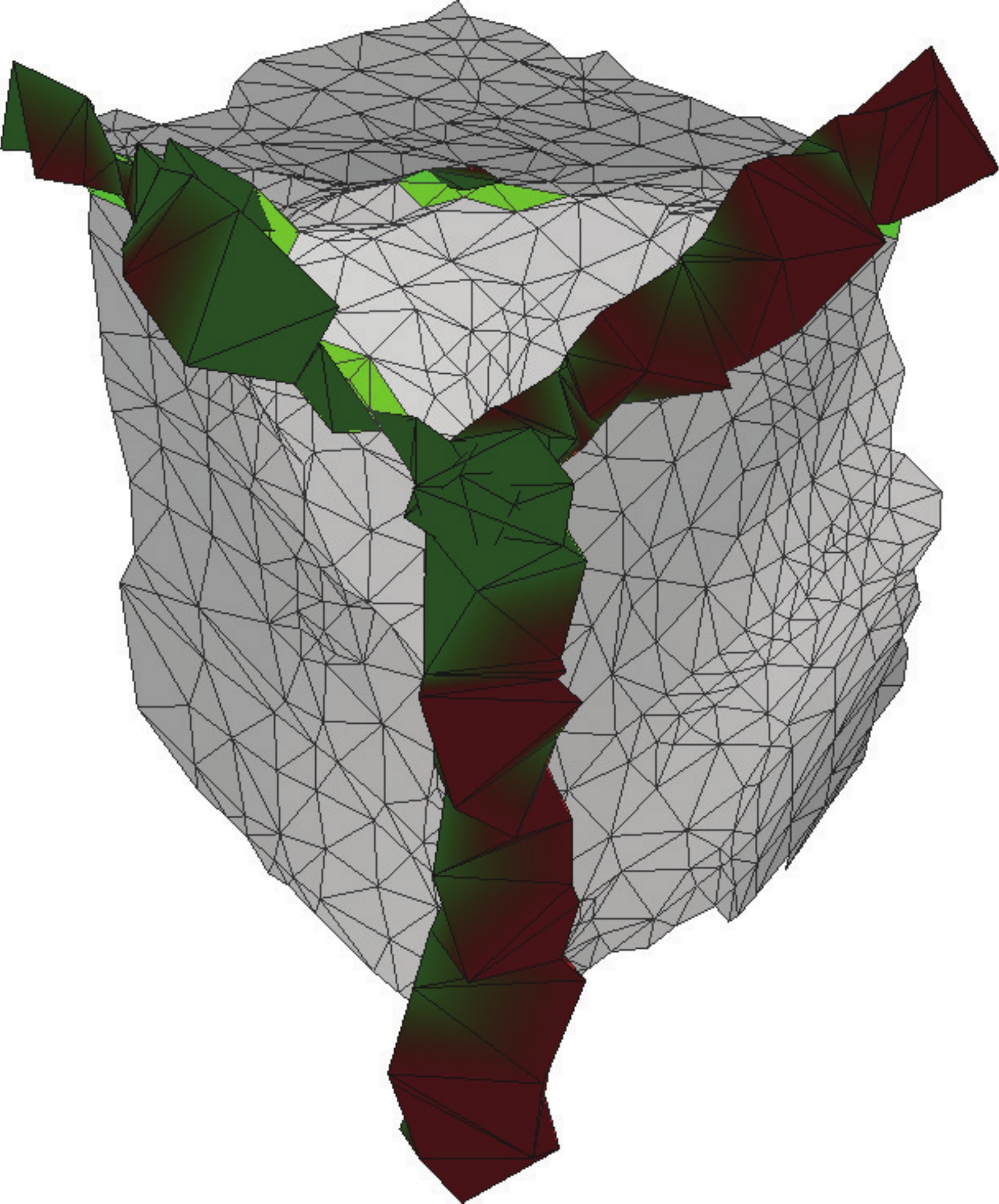}}
\end{minipage}%
\begin{minipage}[b]{.33\linewidth}
    \centering
    \subfloat{\includegraphics[width=0.9\linewidth]{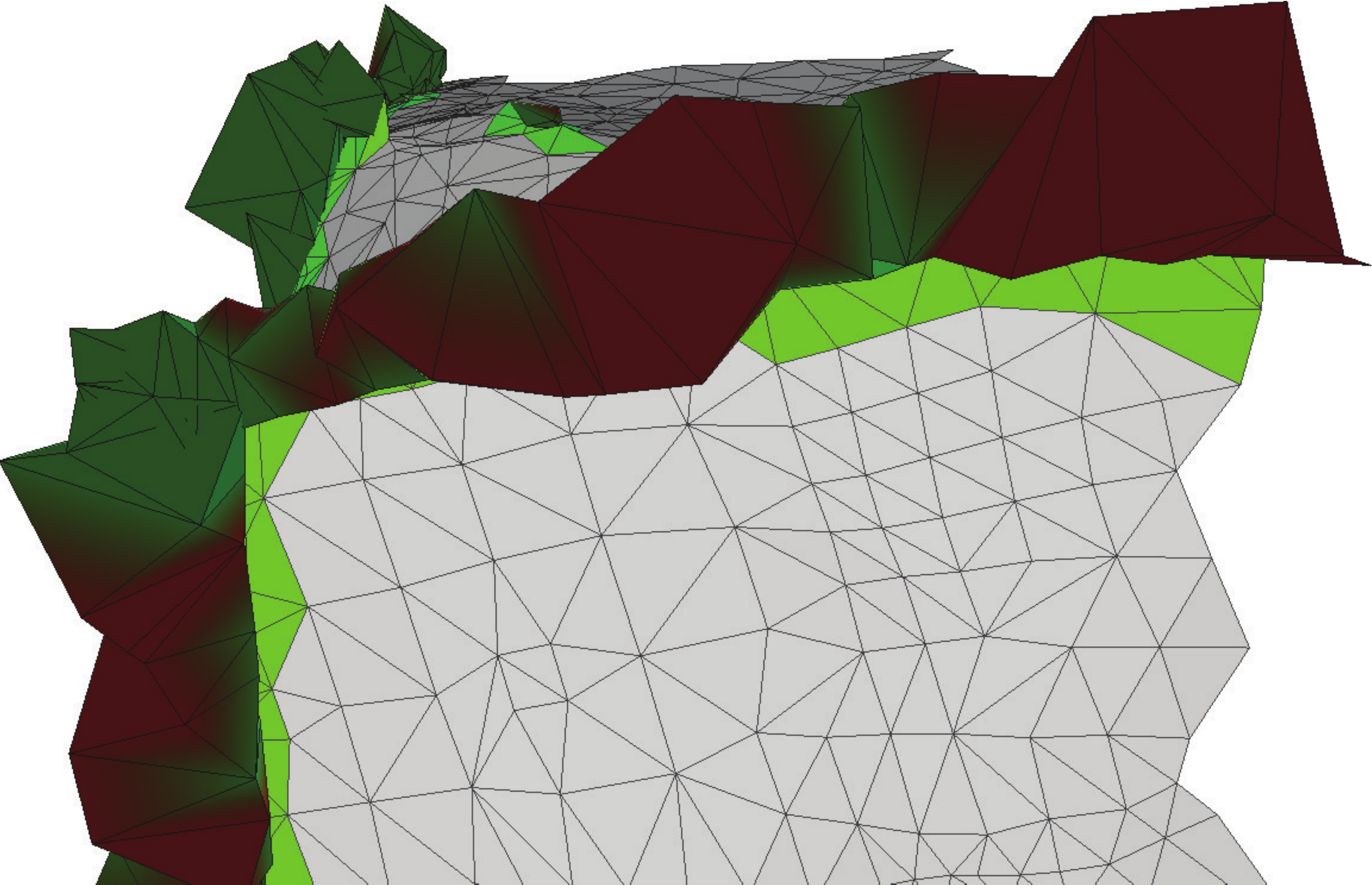}}
    \vfill
    \subfloat{\includegraphics[width=0.9\linewidth]{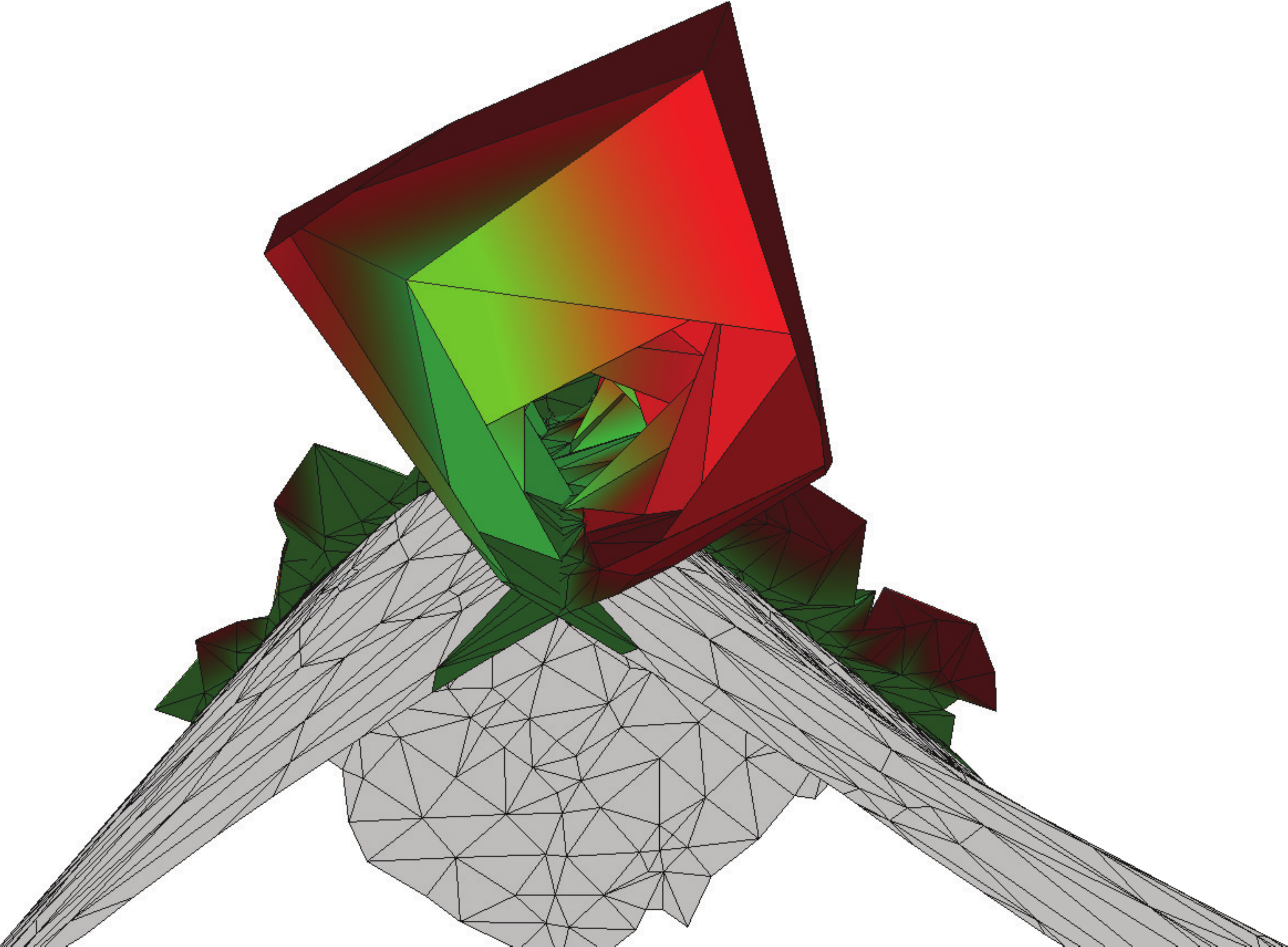}}
\end{minipage}

\caption{\textbf{Left}: Sample rendering of a 3D model that exhibits mesh foldings. \textbf{Right-Top}: Lengthwise detail of the 3D model's side. \textbf{Right-Bottom}: Section of a mesh folding (detail). Color-coding indicates the different parts of the folding: green for the intersecting faces and red for the inward oriented faces. gray faces are part of the unfolded regions.}
\label{fig:folding_example}
\end{figure*}

%%%TODO

The contributions of this paper are novel algorithmic methods that efficiently:

\begin{enumerate}
  \item identify mesh surface foldings. The proposed method detects intersecting faces on the 3D mesh surface as part of the foldings. These faces define one of the key regions of the foldings.

  \item segment mesh surface into distinct regions for both normal and folding surface areas. The separation is achieved via the removal of the previously detected intersecting faces as well as the protruding faces. The latter define a set of faces that falsely connect different (i.e. normal and inward oriented faces) regions only by a limited number of edges. These are successfully identified and removed as well, by the proposed methodology.

  \item remove surface foldings by identifying the inward oriented faces' regions. These regions are robustly identified by the proposed methodology via measurement, on a per region basis, of the number of crossings between the corresponding face normals and the region's surface.

  \item mesh repair which is achieved by reconstructing the gaps created during face removal steps. The mesh reconstruction is guided by the intersection line of the corresponding faces. The proposed methodology aims at reproducing the mesh structure as this is produced after the removal of the mesh singularities previously detected.
\end{enumerate}

The choice of face removal against local remeshing is done to better simulate processes like erosion and degradation of 3D models. These would results in `loss' of mesh structure rather than a differentiation of it. Hence a `removal' process is considered a better fit. However, since the preservation of the original structure of the 3D model is critical, one of the aforementioned novelties of the proposed method is that the gap filling is guided by the intersection line of the corresponding faces removed.

To facilitate comparative studies, all 3 versions (original object, folded version after offsetting operator, repaired version) of the models used in the evaluation are provided for download. Unfortunately, since the majority of previously published methodologies are evaluated against proprietary 3D model sets that have not been made publicly available, a comparative study is not feasible. However, this should be possible for future works.

The remainder of this paper is organized as follows: Section~\ref{sec:problem_description} describes in detail the problem that this paper addresses and Section~\ref{sec:related_work} discusses state-of-the-art works as these have been proposed for the specific problem of mesh foldings. Both reactive and proactive solutions are presented. Section~\ref{sec:proposed_method} details the proposed method. Section~\ref{sec:results} presents the experimental evaluation of the proposed method while an evaluation of its computational complexity is also given. Finally, Section~\ref{sec:conclusions} summarizes and concludes. % as well as sets goals for future research.

\section{The Problem of Mesh Folding}
\label{sec:problem_description}

``Correctness'' of polygonal models is crucial for the model's physical consistency and its suitability for further processing or visualization. The most common types of defects that occur in typical mesh object models are: holes and isles, singular vertices, gaps and overlaps, complex edges, inconsistent normal orientations, and self-intersections \cite{Botch:Book:2010,Ju:JCST:2009}.

Self-intersections are often caused by certain types of mesh deformations \cite{Botch:Book:2010}. The transformation of a given surface $S=\{\mathbf{p}_1,\mathbf{p}_2,\ldots,\mathbf{p}_n\}$ to a deformed surface $S^\prime=\{\mathbf{p}^\prime_1,\mathbf{p}^\prime_2,\ldots,\mathbf{p}^\prime_n\}$, with $\mathbf{p}_i,~\mathbf{p}^\prime_j \in \mathbb{R}^3$, can be described by a displacement function $\mathbf{d}$ that associates to each point $\mathbf{p}_i \in S$ a displacement vector $\mathbf{\hat{d}}_i = \mathbf{d}(\mathbf{p}_i)$, thus mapping the given surface $S$ to its deformed version $S^\prime$:

$S^\prime := \{ \mathbf{p} + \mathbf{d}(\mathbf{p}) \mid \mathbf{p} \in S \}~.$

Mesh foldings can then be exhibited when extreme displacement values are applied on dense surface meshes.

A mesh folding has a distinctive morphology that essentially defines the particular problem. The \emph{folded} surface region is comprised of two directly connected parts: the \emph{inward oriented face} part, whose normal vectors' direction is (although not inverted) oriented \emph{towards} the mesh surface and the \emph{intersecting} part, whose faces intersect with other faces. The \emph{intersecting} part is adjacent to the \emph{unfolded} region of the mesh surface (see Fig.~\ref{fig:folding_example}).

Mesh folding is a very common problem in mesh offsetting procedures and is not adequately tackled in existing literature.

\section{Related Work}
\label{sec:related_work}

%Related work can be found in:
%\cite{Theoharis:Book:2008},
%\cite{Moeller:JGT:1997},
%\cite{Moeller:Book:1999},
%\cite{Botch:Book:2010},
%\cite{Ju:JCST:2009},
%\cite{Haralick:TPAMI:1987},
%\cite{Gueziec:TVCG:2001},
%\cite{Nooruddin:TVCG:2003},
%\cite{Campen:CGF:2010},
%\cite{Liu:TASE:2011},
%\cite{Ju:TOG:2004},
%\cite{Jung:CADA:2004},
%\cite{Kim:CADA:2004},
%\cite{Malosio:CADA:2009},
%\cite{Chen:MISC:NoYr}.

Although much research has been conducted in the area of model repair, the specific problem of detecting and removing foldings from mesh objects using mesh-oriented approaches is rather understudied (see the surveys \cite{Botch:Book:2010} and \cite{Ju:JCST:2009}).

Most model repair algorithms can roughly be classified as \emph{mesh-based} or \emph{volume-based}. Mesh-based algorithms operate directly on the input mesh trying to explicitly identify and resolve defects on the surface. Volume-based algorithms convert the input model into an intermediate volumetric representation which is used for processing and from which the output model is derived, using Marching Cubes \cite{Lorensen:SIGGRAPH:1987}, Dual Contouring \cite{Ju:TOG:2002}, or other mesh extraction algorithms.

Mesh-based repair algorithms only minimally perturb the input model and are able to preserve the polygonal mesh structure, at least in areas that are not in the direct vicinity of the defects. In the case of intersections or large overlaps, these defects cannot be resolved robustly mainly due to numerical inaccuracies. Furthermore, these algorithms introduce a small number of additional triangles in some cases.

On the other hand, by their very nature, volume-based representations do not allow for defects like intersections, holes, gaps, overlaps or inconsistent normal orientations and thus such problems can be solved by appropriately converting a surface model into its volumetric representation. On the downside, the conversion to and from a voxelized model necessitates resampling, which introduces aliasing defects, destroys the structure and connectivity of the input model, introduces numerical approximation errors and is quite memory intensive.

%M\"{o}ller \cite{Moeller:JGT:1997,Moeller:Book:1999} (1997) introduced the so called \emph{interval overlap method}, the first
%method for determining if two triangles intersect. Although a fast intersection test method, it suffers from robustness problems
%when the triangles are nearly co-planar or an edge is nearly co-planar to the other triangle. To resolve this issue a user defined
%approximation threshold was introduced, making the method parameter depended.

Nooruddin and Turk \cite{Nooruddin:TVCG:2003} introduced one of the first volume-based techniques to repair arbitrary models that contain gaps, overlaps and intersections. Additionally, they employed morphological operators to resolve topological defects like holes and handles, resulting in manifold output models. Morphological operators, such as \emph{dilation} and \emph{erosion}, were first introduced in image processing by Haralick \etal \cite{Haralick:TPAMI:1987}.

Ju \cite{Ju:TOG:2004} presented a method for repairing arbitrary polygonal models by producing a closed surface, i.e., a surface that partitions space into disjoint internal and external volumes. Their approach takes a polygonal model, constructs an intermediate volume grid employing octrees and generates the output surface by utilizing the marching cubes or the dual contouring algorithm. Sharp features in the original model are preserved faithfully in the output.

Campen and Kobbelt \cite{Campen:CGF:2010} presented operators that modify the topology of polygonal meshes at intersections and self-intersections for mesh repair. The vertex-based input mesh is initially transformed into a plane-based representation. Using a triangle-triangle intersection test, self-intersection regions are detected. Each intersection region is then covered by a set of small volume cells using an adaptive refinement approach of the octree and processing is performed locally within each cell. The repaired object is the outcome of boolean operations between the extracted oriented components, applying an outer hull approach. Satisfactory visual results repairing self-intersections on 3D models were presented. %, processing $6-12~K$ faces in $1.2-1.9~sec$.

Gu\'{e}ziec \etal \cite{Gueziec:TVCG:2001} described a technique that removes topological singularities from polygonal surfaces,
converting a non-manifold surface to a manifold one. Their algorithm consists of two high level operations: \emph{cutting} and \emph{stitching}. During cutting, isolated vertices and edges shared by at least three different faces are marked and the surface is broken down to new surfaces. During stitching, boundary edges that were created during the cutting operation are pinched together and edges that are close to one another are snapped, yielding a manifold surface. Their method actually eliminates surface singularities such as isolated vertices and multi-shared edges, but cannot address the problem of surface self-folding.

Jung \etal \cite{Jung:CADA:2004} proposed an algorithm to remove self-intersections from the raw offset triangular mesh. The method's pipeline consists of finding and deleting degenerate triangles, computing the self-intersection segments, and detecting valid triangles, utilizing a region growing procedure starting from a seed triangle. Finally, triangle splitting, triangle trimming and stitching the self-intersection segments concludes the procedure. %Processing time ranges for $15-256~K$ faces at $2.5-53.9~sec$.
The method was tested on outward offset meshes with good results. However, since the valid triangles detection process is based on seeding from a valid triangle, the automatic detection of such seeds cannot be guaranteed on surface models exhibiting mesh foldings, especially when the deformations are of big extent. Furthermore, since the method relies on local characteristics of the mesh structure (i.e. does not take into consideration the orientations of mesh surface regions), it could possibly produce wrong results, again especially in the case of large deformations, where the \emph{folded} regions are (almost) equal to the \emph{unfolded} regions. %****TT: WHY DO THESE HOLD FOR LARGE DEFORMATIONS? I DO NOT UNDERSTAND}

We next present methods for mesh offsetting, aiming to avoid self-intersections by construction.

Kim \etal \cite{Kim:CADA:2004} introduced a method for offsetting triangular meshes by moving vertices along multiple normal vectors at each vertex. These vectors are computed by the normal vectors of the faces surrounding the vertex. Offsetting with the multiple normal vectors of a vertex does not create a gap or overlap at the smooth edges, making the mesh size uniform and the computation time short. However, this method can also create self-intersections that have to be removed.
%The vertices are moved along the multiple normal directions of a vertex \hl{TT: HOW CAN MULTIPLE VERTICES BE MOVED ALONG THE MULTIPLE NORMAL DIRECTIONS OF A SINGLE VERTEX? UNCLEAR}, and the gaps at sharp edges are filled by a blending mesh. This method, although it creates a smooth offset surface, can also create self-intersections that have to be removed.

Malosio \etal \cite{Malosio:CADA:2009} described a geometric algorithm to offset triangular mesh surfaces, transforming the original
geometry to a new smooth surface model. The method calculates a corrected direction and magnitude for the displacement vector that
represents the offsetting of each vertex. Although the method gives smooth results to the sharp edge offset, it has some drawbacks since
it exhibits distortions, such as self-intersections in some cases.

Liu and Wang \cite{Liu:TASE:2011} presented an intersection-free surface offset generation approach for free-form triangular meshes
which preserves sharp features. Their method for offset surface generation can be classified as a volumetric approach. Firstly, the given
model is sampled in a uniform volumetric grid to construct a signed distance field; next, the intersections between the grid edges
and the offset surface are computed and finally, an intersection-free dual contouring algorithm is introduced to extract the offset
surface mesh.

Chen \etal \cite{Chen:MISC:NoYr} proposed an approach for computing offsets of solids bounded by triangular meshes. Their approach is
based on a hybrid data structure combining point samples, voxels and continuous surfaces. It samples the boundary, generates surfels
in the vicinity of the model, uses a spatial grid to eliminate candidate samples that are too close to the original surface and restores
the offset boundary through interpolation. The method was tested on outward and inward offset meshes showing visually satisfactory results.

Although methods that avoid self-intersections by construction can generally produce good results, the aim of this paper is to detect and repair mesh foldings after they have occurred on a 3D mesh. As in the case of most applications, the original (\emph{unfolded}) 3D model will not be available.

\section{The Mesh Folding Recognition and Repair Method}
\label{sec:proposed_method}

In this section, the proposed method for the recognition and repair of foldings on the mesh surface of 3D models is described in detail. For the remainder of this paper the proposed method shall be referred to, as \textbf{Fo}lding \textbf{R}ecogntition and \textbf{R}epair \textbf{M}ethod, or \textbf{FoR$^2$M}.

A simple mesh surface representation for both input and output 3D models, is assumed: each mesh surface is described by a set of vertices $V$ (where $V_i$, $i\in1..n$, defines the $i^{th}$ vertex) and a set of triangular faces $F$ (where $F_j$, $j\in1..m$, defines the $j^{th}$ face). Other structures, necessary in specific processing steps (e.g. vertex and face neighborhood connectivity, boundary edges, etc) are computed from the aforementioned input.

\begin{figure}
  \centering
  \includegraphics[width=.8\linewidth]{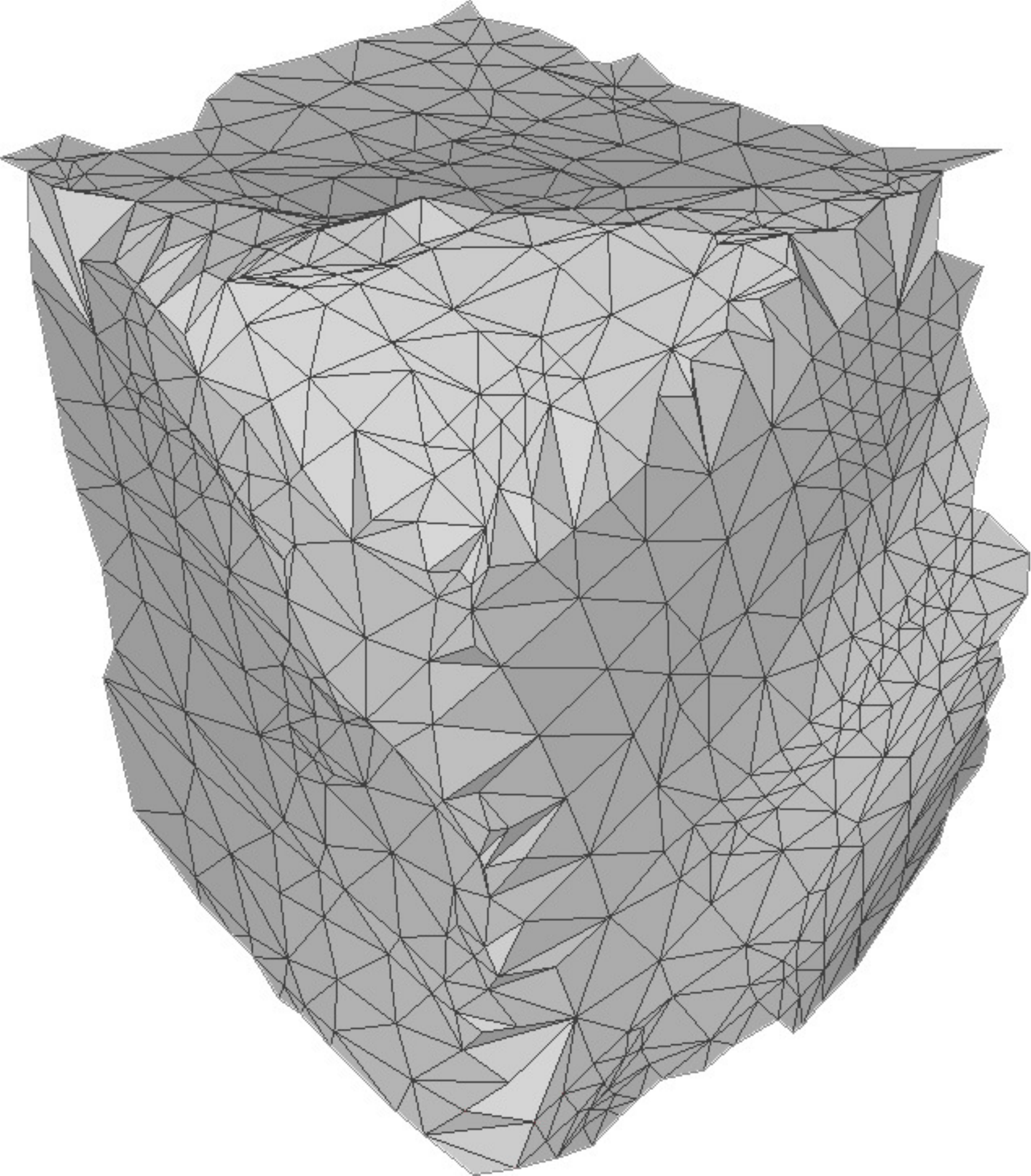}
  \caption{Sample mesh surface exhibiting foldings. \emph{Unfolded} connected components are retained and color labeled. Gaps between the connected components have been reconstructed and filled. The final output mesh surface of the algorithm is illustrated.}
  \label{fig:folding_step_05}
\end{figure}

\textbf{FoR$^2$M} is outlined as follows:

\begin{enumerate}
\setlength\itemsep{1em}

\item\label{list:detection}
\emph{Detection of mesh intersecting faces.} The intersecting faces of the mesh foldings (denoted as set \textit{I}) are detected. Subsequent steps use \textit{I} to effectively distinguish between \emph{folded} and \emph{unfolded} regions of the mesh surface. Intersection lines are also computed. See Fig.~\ref{fig:folding_step_01}.

\item\label{list:removal_lower}
\emph{Removal of intersecting and protruding faces.} The intersecting faces \textit{I} are removed from the mesh surface. This procedure results in the separation of the mesh surface into \emph{folded} and \emph{unfolded} regions. Furthermore, any \emph{protruding} faces, created by the removal procedure of this step, are also removed. These are faces, at least 1/2 of whose adjacent faces are intersecting faces. See Fig.~\ref{fig:folding_step_02}.

\item\label{list:partitioning}
\emph{Mesh partitioning.} In this step, connected components are formed by grouping together neighboring faces of the mesh surface. See Fig.~\ref{fig:folding_step_03}.

\item\label{list:labeling}
\emph{Connected components labeling.} The connected components of the previous step are labeled as \emph{folded} or \emph{unfolded}. Each \emph{folded} connected component is defined by a set of \emph{inward oriented faces} (set \textit{R}). A simple intrinsic feature is used for the classification. See Fig.~\ref{fig:folding_step_03} and Fig.~\ref{fig:folding_step_04}.

\item\label{list:removal_upper}
\emph{Removal of folded connected components.} In this step the connected components of set \textit{R} %the mesh foldings that are composed of \emph{inward oriented faces}
are removed. Furthermore, any connected components of insignificant size, that may have been created during the previous steps, are also removed. See Fig.~\ref{fig:folding_step_04}.

\item\label{list:reconstruction}
\emph{Gap reconstruction.} The intersecting faces \textit{I} of step~\ref{list:detection} are split along the intersection line. The resulting faces that are directly attached to the \emph{unfolded} connected components are used to complete the gaps that (may) have been created by the removal procedures of steps \ref{list:removal_lower} and \ref{list:removal_upper}. See Fig.~\ref{fig:folding_step_05}.

\item\label{list:filling}
\emph{Gap filling.} Finally, any remaining gaps are filled with new triangular faces. These gaps are usually created in areas where more than one intersection lines existed (i.e. multiple intersected triangles). See Fig.~\ref{fig:folding_step_05}.
\end{enumerate}

Steps \ref{list:detection} and \ref{list:removal_lower} constitute the \textbf{Mesh Intersection Processing}. Steps \ref{list:partitioning}, \ref{list:labeling} and \ref{list:removal_upper} constitute the \textbf{Connected Component Processing}, while steps \ref{list:reconstruction} and \ref{list:filling} comprise the \textbf{Mesh Repair} process.

In the following sections, the aforementioned steps are described in detail.

\subsection{Mesh Intersection Processing}
\label{sec:mesh_int_process}

\subsubsection{Detection of Mesh Intersecting Faces}

%In the first step of the proposed method, recognition of the mesh foldings locations on the mesh surface is performed.
A mesh folding is composed of two distinct sets of faces: the \emph{intersecting} set \textit{I}, comprised of faces that intersect with each other and the \emph{inward oriented faces} set \textit{R}, comprised of faces whose normal vectors are oriented \emph{towards} the mesh surface (although not inverted).

% OK TODO: *** describe this in a previous section ^^

\begin{figure}[t]
  \centering
  \includegraphics[width=.8\linewidth]{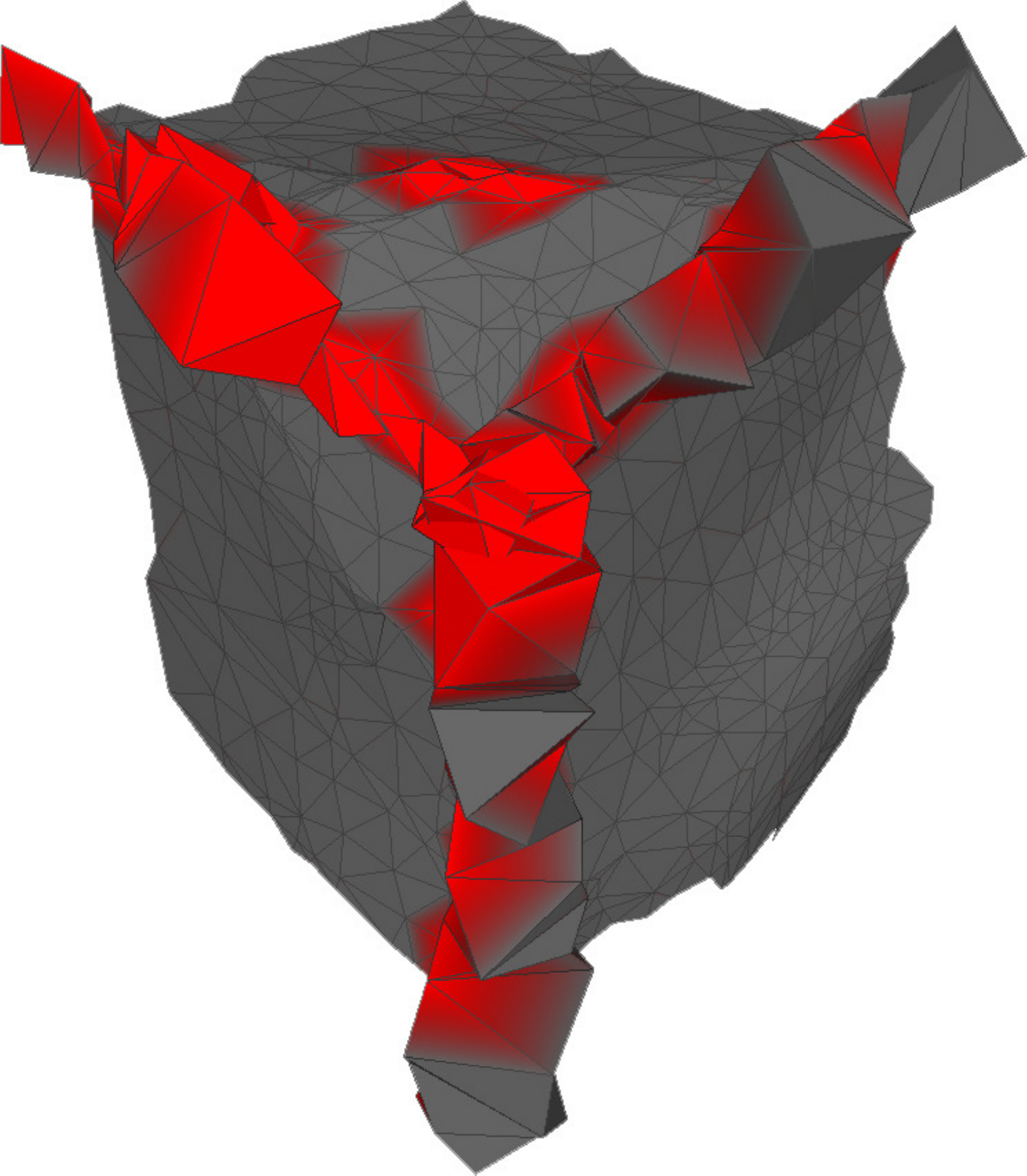}
  \caption{Sample mesh surface exhibiting foldings. Intersecting faces are indicated.}
  \label{fig:folding_step_01}
\end{figure}

The intersecting faces \textit{I} are used to efficiently locate the mesh foldings (See Fig.~\ref{fig:folding_step_01}). In previous works, many algorithms that solve the problem of detecting mesh surface intersections have been proposed, a well-known such algorithm being the triangle-triangle intersection test proposed by M\"{o}ller \cite{Moller:1997:FTI:272317.272320}. During our experimentations we have discovered that although M\"{o}ller's algorithm is applicable to the specific problem, it lacks efficiency in meshes that are predominantly comprised of large near-planar surfaces.

M\"{o}ller's algorithm initially performs a triangle-triangle intersection assessment, where each reference triangle $F_r$ (with vertices $V_{r,0}, V_{r,1}, V_{r,2}$, where $r~\in~1..m$) is tested against every other (test) triangle $F_t$ (with vertices $V_{t,0}, V_{t,1}, V_{t,2}$, where $t~\in~1..m, t~\neq~r$). This step attempts to eliminate a large number of test triangles, by testing if all the vertices of a test triangle lie only on the same side of the plane of a reference triangle ($\pi_r : N_r \cdot X + d_r = 0$, where $X$ is any point in the plane), where:

\begin{align*}
  &N_r = (V_{r,1} - V_{r,0}) \times (V_{r,2} - V_{r,0})\\
  &d_r = -N_r \cdot V_{r,0}
\end{align*}

The signed distances from the vertices of $F_t$ to $\pi_r$ (multiplied by a constant $N_r \cdot N_r$) are computed by inserting the vertices into the plane equation:

\begin{align*}
  &d_{V_{t,i}} = N_r \cdot V_{t,i} + d_r  \intertext{where } &i = 0,1,2
\end{align*}

If all $d_{V_{t,i}}\neq0, i=0,1,2$ (no point lies on the plane) and all have the same sign then $F_t$ lies on one side of plane $\pi_r$ and the overlap is rejected.

%However, in the case of planar or near-planar surfaces most of the test triangles will pass this test since, although not intersecting with the reference triangle, their vertices lie on either side of the reference plane (i.e. do not have the same sign).

For planar or quasi-planar mesh surfaces, large numbers of candidate triangle pairs will not be discarded by M\"{o}ller's algorithm. Subsequent and more time consuming pairwise triangle-triangle intersection tests will need to assess these pairs. See Sec. \ref{sec:comp_comp} for performance figures. To improve performance, an initial intersection assessment based on the axis-aligned bounding volume of each triangle is proposed.

\begin{figure}[t]
  \captionsetup[subfloat]{labelformat=empty}
  \centering
  \subfloat[Unresolved]{\includegraphics[width=0.65\linewidth]{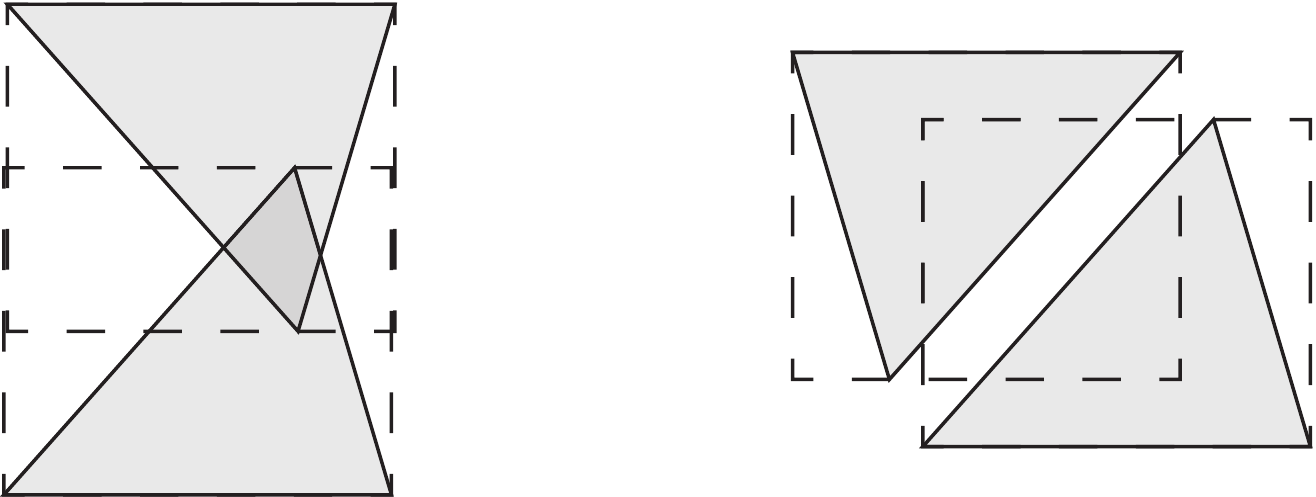}} \\
  \subfloat[Non-intersecting]{\includegraphics[width=0.9\linewidth]{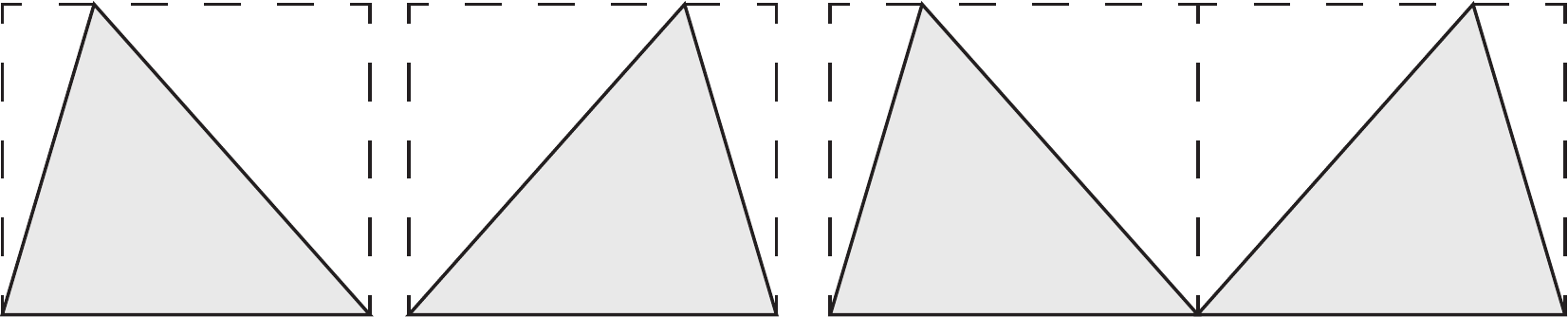}}

  \caption{Triangle bounding volume intersection possibilities. \textbf{Top-left}: overlapping bounding volumes, intersecting triangles. \textbf{Top-right}: overlapping bounding volumes, non-intersecting triangles. \textbf{Bottom-left}: non-overlapping bounding volumes, non-intersecting triangles. \textbf{Bottom-right}: overlapping bounding volumes, non-intersecting triangles (adjacent triangles).}
  \label{fig:triangle_intersections}
\end{figure}

The bounding volume of the test triangle $F_t$ is examined for lying inside the bounding volume of the reference triangle $F_r$ %(Eq.~\ref{eq:dAABB}).
If it does not %and this is also true for the inverse test (i.e. the reference triangle $F_r$ does not lie inside the bounding volume of test triangle $F_t$)
then the two triangles do not intersect and they can be safely excluded from further testing (see Fig.~\ref{fig:triangle_intersections} bottom-left).

%\begin{align*}
%  &dmin(F,d)=min(V_0^d,V_1^d,V_2^d), \\
%  &dmax(F,d)=max(V_0^d,V_1^d,V_2^d),
%  %\intertext{where } &d=X,Y,Z\\
%\end{align*}
%\begin{align}
%\begin{split}
%\label{eq:dAABB}
%    dmin(F,d)= &min(V_0^d,V_1^d,V_2^d), \\
%    dmax(F,d)= &max(V_0^d,V_1^d,V_2^d)\\\\
%  dAABB(F_r, F_t) = &dmin(F_r,d) \leq dmax(F_t,d) \wedge \\
%                    &dmax(F_r,d) \geq dmin(F_t,d) \wedge ...\\
%                    \forall d \in \{X,Y,Z\}
%\end{split}
%\end{align}

%\begin{align}
%\begin{split}
%\label{eq:dAABB}
%    &Collision(F_r, F_t) = \\
%    &(F_r^{min(X)} \leq F_t^{max(X)} \wedge F_r^{max(X)} \geq F_t^{min(X)}) \wedge\\
%    &(F_r^{min(Y)} \leq F_t^{max(Y)} \wedge F_r^{max(Y)} \geq F_t^{min(Y)}) \wedge\\
%    &(F_r^{min(Z)} \leq F_t^{max(Z)} \wedge F_r^{max(Z)} \geq F_t^{min(Z)})\\
%\end{split}
%\end{align}

A large number of triangle pairs that share one or two vertices (neighbors) will pass the above test but should not be considered as intersecting. Such triangles will be discarded in the subsequent pairwise intersection test (see Fig.~\ref{fig:triangle_intersections} bottom-right).

% OK TODO: *** use image and show this example

 The intersections of remaining triangle pairs are then precisely computed using the basic line-triangle intersection test. The edges of each test triangle $F_t$ (i.e. $(V_{t,0},V_{t,1})$, $(V_{t,0},V_{t,2})$ and $(V_{t,1},V_{t,2})$) are tested for intersection against reference triangle $F_r$ and vice versa. In the case of intersection the line of intersection is also computed and stored.

The line-triangle intersection test is a modified version of the method proposed by M\"{o}ller and Trumbore in \cite{Moller:1997:FMS:272313.272315}, taking into consideration lines, rays and line segments as well.

A line $p(t)$ specified by two points $p_1, p_2$ can be defined using its parametric equation:

\begin{equation*}
  p(t)=p_1+t(p_2-p_1)
\end{equation*}

a triangle $F_j$ with vertices $V_{j,0},V_{j,1},V_{j,2}$, can be defined by its parametric equation in terms of its barycentric coordinates:

\begin{equation*}
  F_j(u_1,u_2)=(1-u_1-u_2)V_{j,0}+u_1V_{j,1}+u_2V_{j,2}
\end{equation*}

To test for the intersection of the two objects, the above equations are combined:

\begin{align*}
  &(1-u_1-u_2)V_{j,0}+u_1V_{j,1}+u_2V_{j,2}=p_1+t(p_2-p_1)\Leftrightarrow\\
  &(p_1-p_2)t+(V_{j,1}-V_{j,0})u_1+(V_{j,2}-V_{j,0})u_2=p_1-V_{j,0}
\end{align*}

This constitutes a $3 \times 3$ linear system in $t$, $u_1$ and $u_2$. The solution of this linear system gives the barycentric coordinates of the intersection point with respect to the triangle. The point lies inside the triangle if $0 \leq u_1, u_2 \leq 1$ (which implies $0 \leq 1 - u_1 - u_2 \leq 1$ as well). Furthermore, the parametric value $t$ of the intersection point along the line is given. This is checked for inclusion in the appropriate interval if a ray $(t > 0)$ or a line segment $(t \ge 0 \wedge t \leq 1)$ is being considered.

\subsubsection{Removal of intersecting and protruding faces}

\begin{figure}[t]
  \centering
  \includegraphics[width=.8\linewidth]{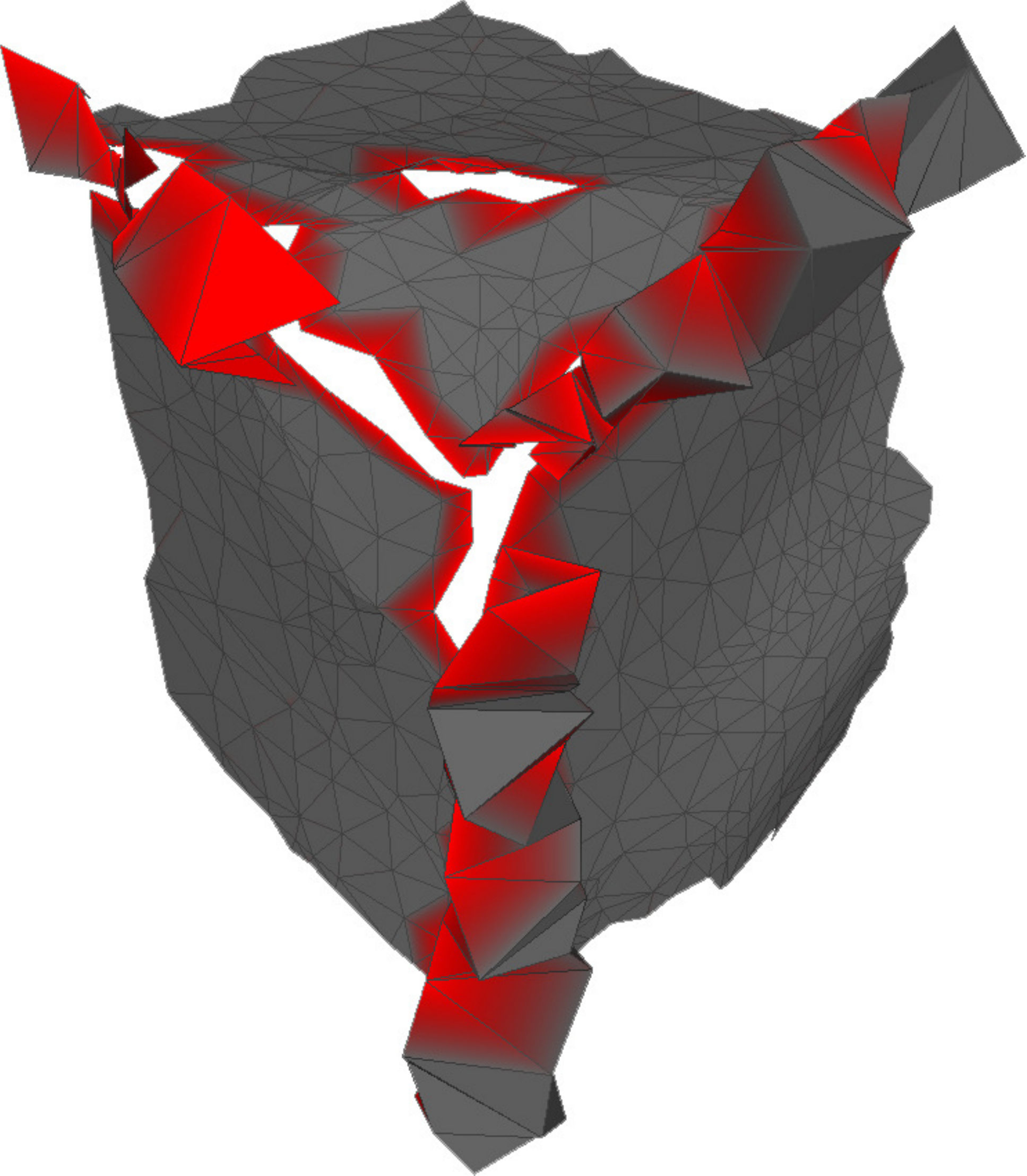}
  \caption{Sample mesh surface exhibiting foldings. Intersecting faces are removed. Edges adjacent to the intersecting faces are indicated.}
  \label{fig:folding_step_02}
\end{figure}

The set of \emph{intersecting} faces \textit{I} is next removed from the mesh surface (See Fig.~\ref{fig:folding_step_02}), thus accomplishing a separation between the remaining \emph{folded} and \emph{unfolded} regions.

We have experimentally found that in order to elevate the quality of the final output model, \emph{protruding} faces also need to be removed. %These protruding faces, although not recognized as intersecting, can be safely categorized as such, since $\geq50\%$ of their adjacent faces are intersecting.
The criteria that a face needs to meet in order to be characterized as protruding are the following (all required):

\begin{itemize}
\item
  A face must not have been identified as intersecting (i.e. $\not\in$~ \textit{I}).
\item
  no. of adjacent faces $> 1$
\item
  no. of intersecting adjacent faces $\ge 1$
\item
  (no. of adjacent faces) - (no. of intersecting adjacent faces) $< 2$
\end{itemize}

%\begin{itemize}
%\item
%  A face must not have been identified as intersecting (i.e. $\not\in$~ \textit{I}).
%\item
%  adjacent faces $\in {2,3}$
%\item
%  intersecting adjacent faces $\in {1,2,3}$
%\item
%  (adjacent faces) - (intersecting adjacent faces) $< 2 =>$\\
%  (adjacent faces) = 2, intersecting adjacent faces $\in {1,2}$\\
%  (adjacent faces) = 3, intersecting adjacent faces $\in {1,2,3}$
%\end{itemize}

Conceptually, the aforementioned criteria imply that a protruding face has two or more adjacent faces and at least half of them are intersecting (and have thus been removed). Note that a face that has exactly one adjacent face, and this is intersecting, defines a single-face connected component which, due to its size, will be removed at a later step.

\subsection{Connected Components Processing}

\subsubsection{Mesh Partitioning}

\begin{figure}
  \centering
  \includegraphics[width=.8\linewidth]{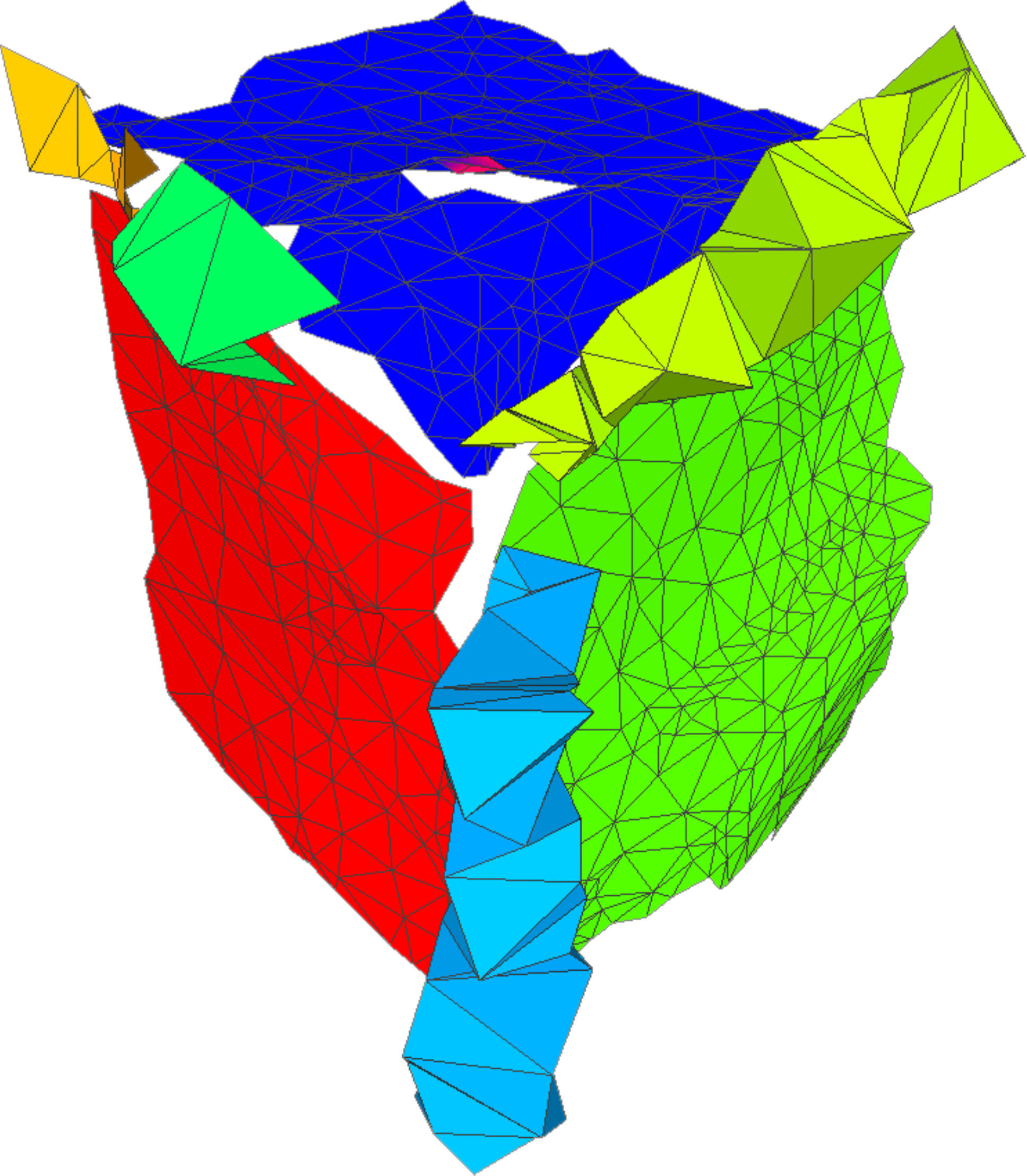}
  \caption{Sample mesh surface exhibiting foldings. Connected components are color labeled. Illustrated are both \emph{folded} and \emph{unfolded} regions.}
  \label{fig:folding_step_03}
\end{figure}

Following the removal of the \emph{intersecting} part of the foldings, the remaining faces need to be characterized as to whether they are part of the \emph{folded} or \emph{unfolded} mesh surface (See Fig.~\ref{fig:folding_step_03}). %In the latter case, these faces need to be removed.

A first step toward this goal is to group all edge-connected faces together, forming connected components. The partitioning scheme employed is a methodology for finding strongly connected components, based on edge adjacency connectivity of the faces \cite{journals/siamcomp/Tarjan72,Aho:1983:DSA:577958}.

It should be noted that, due to the mesh surface complexity of the 3D models, the classic recursive connected component partitioning methods fail to perform on most medium to large scale mesh surfaces. A non-recursive version has thus been employed. %%%%(See Algorithm~\ref{alg:non_rec_conn_comp}). Note that the algorithm does not take into consideration non-manifold mesh surfaces, as these singularities have been eliminated during the previous step.

%\subsection{Connected Component Labeling}
\subsubsection{Connected Components Labeling}

\begin{figure}[t]
  \centering
  \includegraphics[width=.8\linewidth]{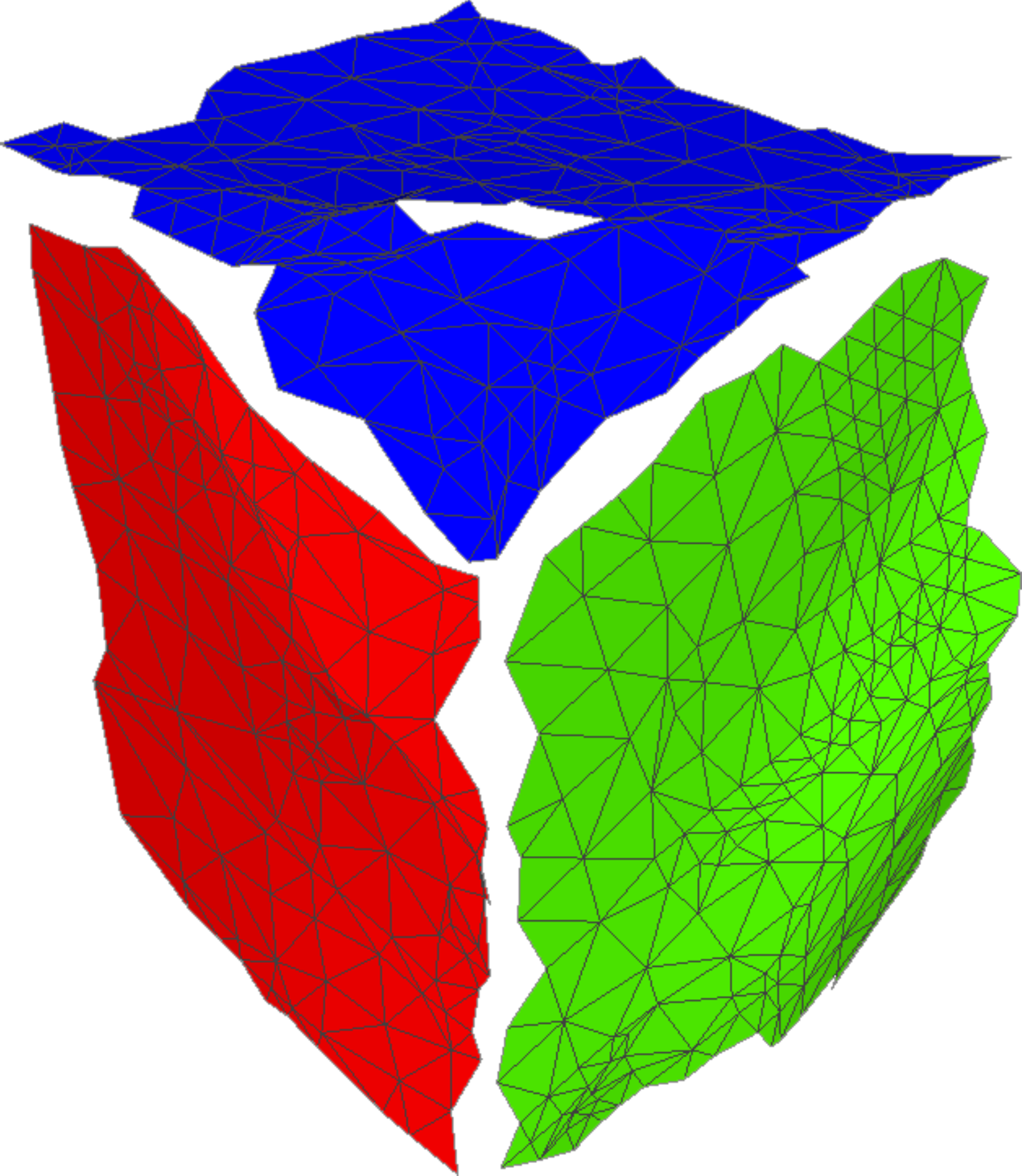}
  \caption{Sample mesh surface exhibiting foldings. Connected components comprised of \emph{inward oriented faces} have been removed. \emph{Unfolded} connected components are retained and color labeled.}
  \label{fig:folding_step_04}
\end{figure}

The remaining surface of the foldings is comprised of the set of \emph{inward oriented faces} (\textit{R}), which have their normal vectors' orientation altered, so that they point \emph{towards} the mesh surface. This feature will be exploited in order to determine which connected components should be recognized as parts of the foldings (set \textit{R}) and should thus be removed (See Fig.~\ref{fig:folding_step_03} and Fig.~\ref{fig:folding_step_04}).

In an \emph{unfolded} connected component, the face normal vectors are oriented \emph{outwards} and should thus intersect with the mesh surface an \textbf{even} number of times (including zero). An exception to this rule is the case of mesh surfaces that have holes.

Conversely, in a \emph{folded} connected component, the face normal vectors, being oriented \emph{inwards}, should intersect with the mesh surface an \textbf{odd} number of times.

%In a \emph{folded} connected component, the face normal vectors, oriented \emph{inwards}, intersect with the mesh surface an \textbf{odd} number of times. This is true, since a face normal vector has its origin on the mesh surface and is directed \textbf{towards} it. In case the normal vector intersects with the mesh surface, the number of intersections \emph{entering} the mesh surface would be one less than the number of intersections \emph{exiting}. Again, a exception to the aforementioned rule is the case of mesh surfaces that have holes.

Therefore, to label a connected component as \emph{unfolded} or \emph{folded}, a check on the number of intersections of its face normal vectors and the entire 3D model is performed. The modified line-triangle intersection test by M\"{o}ller and Trumbore \cite{Moller:1997:FMS:272313.272315} is employed for this purpose. For each face, the normal vector emanating from its centroid is tested for intersection against the remaining faces of the same connected component. To compensate for possible holes and other minor mesh defects, the following rule is employed: If 20\% or less of a connected component's face normal vectors have \textbf{odd} number of intersections then this connected component is labeled as \emph{unfolded}, otherwise is labeled as \emph{folded} and should be removed. The aforementioned threshold has been experimentally determined.%found to be efficient in terms of recognition accuracy.

It should be noted that computing the intersection test on a per connected component basis, instead of the complete 3D model mesh surface, significantly reduces the total computational time, since it is a quadratic time algorithm applied on partitioned input sizes. %whereas the results produced are identical.

%\subsection{\emph{Irregular} Surface Areas Removal}
\subsubsection{Removal of \emph{Folded} Connected Components}

In this step, the \emph{folded} connected components that contain \emph{inward oriented faces} (\textit{R}), as well as connected components of \emph{insignificant size}, are removed (See Fig.~\ref{fig:folding_step_04}). The latter are outlier mesh surface regions that (may) have been created during any of the previous removal steps. % NOT TODO: *** have an image showing this!
The insignificant size criterion is $l = 1\%$ of the total number of faces of the mesh surface. To this end, all connected components are sorted based on the number of faces that they are composed of; these numbers are then summed from the beginning of the list until percentage $100-l$ is reached. Any connected components that remain, are discarded. Note that it is possible that no connected components are removed at all. %if all partitions are composed of large number of faces (with respect to the total number of faces of the mesh surface).

After this step, only the \emph{unfolded} mesh surface remains. %The final step of the proposed method will complete the repair of the 3D model by reconstructing the erroneous mesh surface which was removed between the retained connected components.

\subsection{Mesh Repair}

\subsubsection{Gap Reconstruction}

The mesh repair step aims at stitching the gaps created by the face removal steps between the connected components. To do this efficiently, the set of intersecting faces \textit{I} that were removed in \textbf{Mesh Intersection Processing} are utilized.

\begin{figure}
  \centering
  \includegraphics[width=0.7\linewidth]{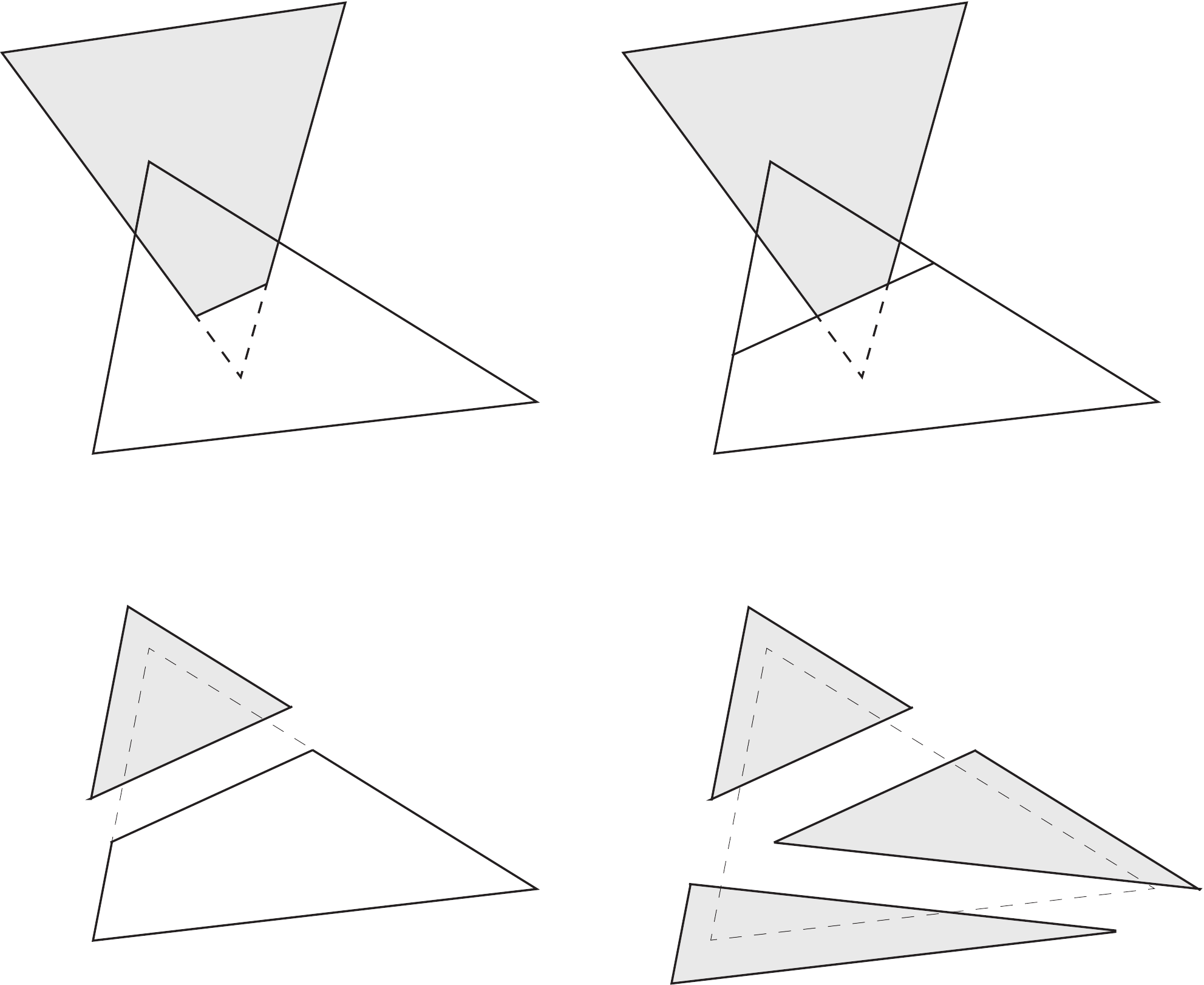}
  \caption{Triangle splitting example. \textbf{Top-left}: an intersected triangle with its intersector. \textbf{Top-right}: the intersection line on the intersected triangle. \textbf{Bottom-left} the single split triangle on one side of the intersection line. \textbf{Bottom-right}: the two split triangles on the other side of the intersection line.}
  \label{fig:triangle_splitting}
\end{figure}

More specifically, for each triangle (referred to as \emph{intersected} triangle) a list is created, containing every other triangle (referred to as \emph{intersector}) that intersects with it. For the first of its intersectors, the intersected triangle is split along the corresponding intersection line. The splitting procedure results into three new triangles which replace the initial intersected triangle (see Fig.~\ref{fig:triangle_splitting}).
One of the three resulting triangles lies on one side of the intersection line and the other two on the other side. To this effect, the edges of the \emph{intersected} triangle, which are crossed by the intersection line are considered. The side that forms a triangle is retained as is and the side that forms a trapezoid is split so that two new triangles are formed.

% OK TODO: *** add image of triangle splitting

%For each of the new intersected triangles, the procedure is recursively repeated using the next intersector in list.
Each of the 3 newly generated triangles is then iteratively considered for intersection with the next intersector in the list. %As a result,
In the end, the initial intersected triangle is split along every intersection line of its intersectors. The iterative intersected triangular face splitting strategy is outlined in Algorithm~1. %***\ref{alg:tri_split}.
Note that the algorithm presented is not optimized and intensionally outlined as iterative in order to require as little as possible memory space (something crucial in cases of large mesh surfaces that have many intersecting triangles).

To perform the reconstruction of the gaps, all intersected triangles are tested for adjacency with the retained connected components of the previous step. Here, two cases exist: (a) a single triangle lies on one side of the intersection line (Fig.~\ref{fig:triangle_splitting}, bottom-left) and (b) two triangles lie together on one side of the intersection line (Fig.~\ref{fig:triangle_splitting}, bottom-right). In case (a) this triangle is solely used for the reconstruction while in (b) both triangles are employed even if they are not both directly adjacent to a connected component.

\begin{figure}
  \centering
  \includegraphics[width=0.7\linewidth]{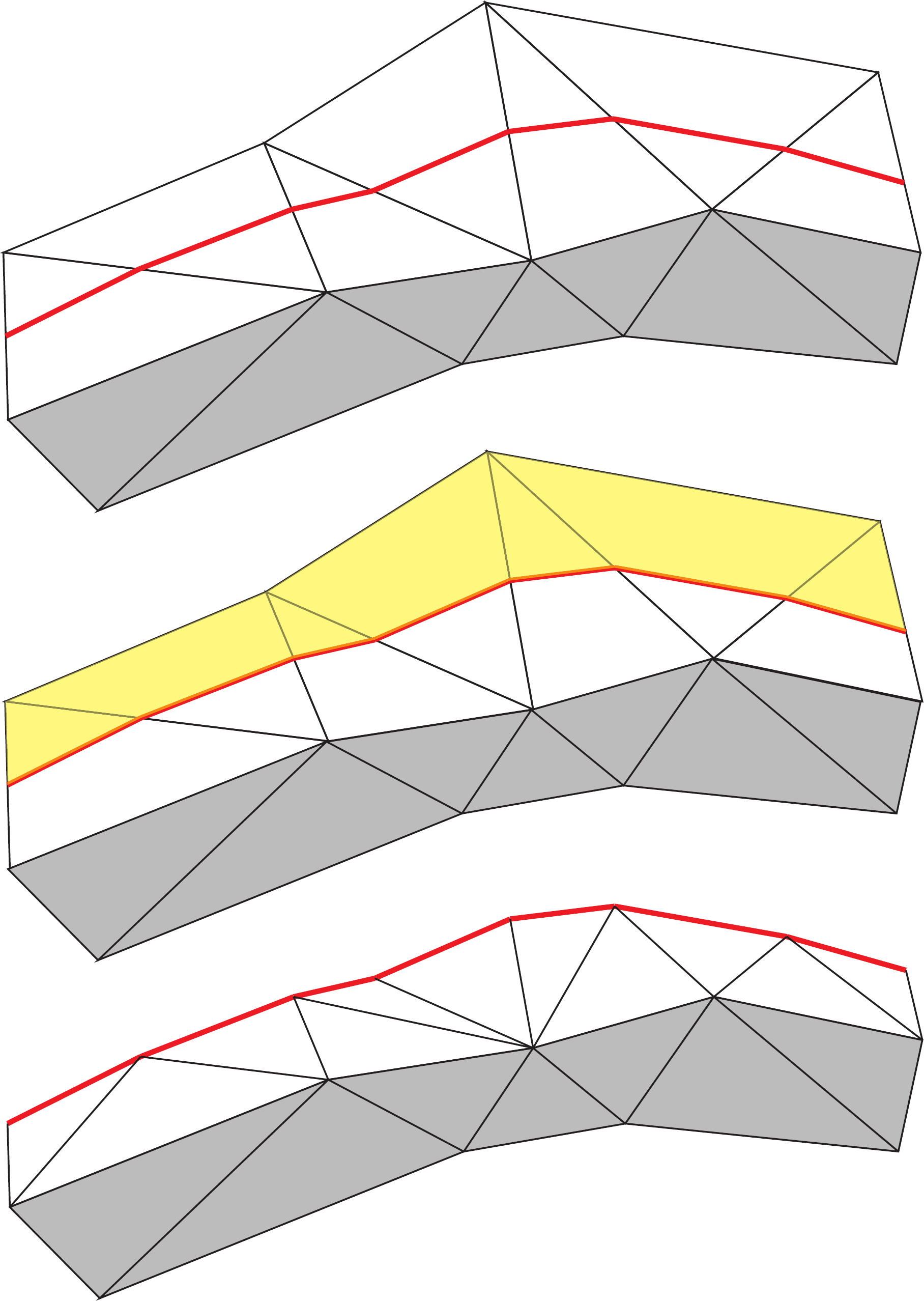}
  \caption{Gap reconstruction example. \textbf{Top}: sample intersecting triangles that are on one edge adjacent to the normal surface (indicated by darker triangles). The intersection line is also indicated. \textbf{Middle}: the indicated part of the intersecting triangles must be removed as it is not adjacent to the normal surface and is located on the side of the intersection line that will not be retained. \textbf{Bottom}: the resulting triangle mesh structure after application of the proposed algorithm. The retained triangle area is remeshed so that the intersection line is exactly traced.}
  \label{fig:gap_stitching}
\end{figure}

The intersection line(s) defines the boundary between the retained mesh surface and the (folded) surface (see Fig.~\ref{fig:gap_stitching}). If a triangle lies on the side of the retained surface, even if not adjacent to the surface, it is retained. The reconstruction faces retained are denoted as \textit{$I_{R}$} and they are a subset of \textit{I}, where some of the faces may be split. The gap reconstruction algorithm is outlined in Algorithm~2. %***\ref{alg:gap_reconstruction}.

\subsubsection{Gap Filling}

%In the final step of the proposed method
A gap filling strategy aims at completing the mesh repair procedure and it is necessary when multiple intersections exist and holes are created between them.

\begin{figure}[ht]
  \centering
  \includegraphics[width=0.7\linewidth]{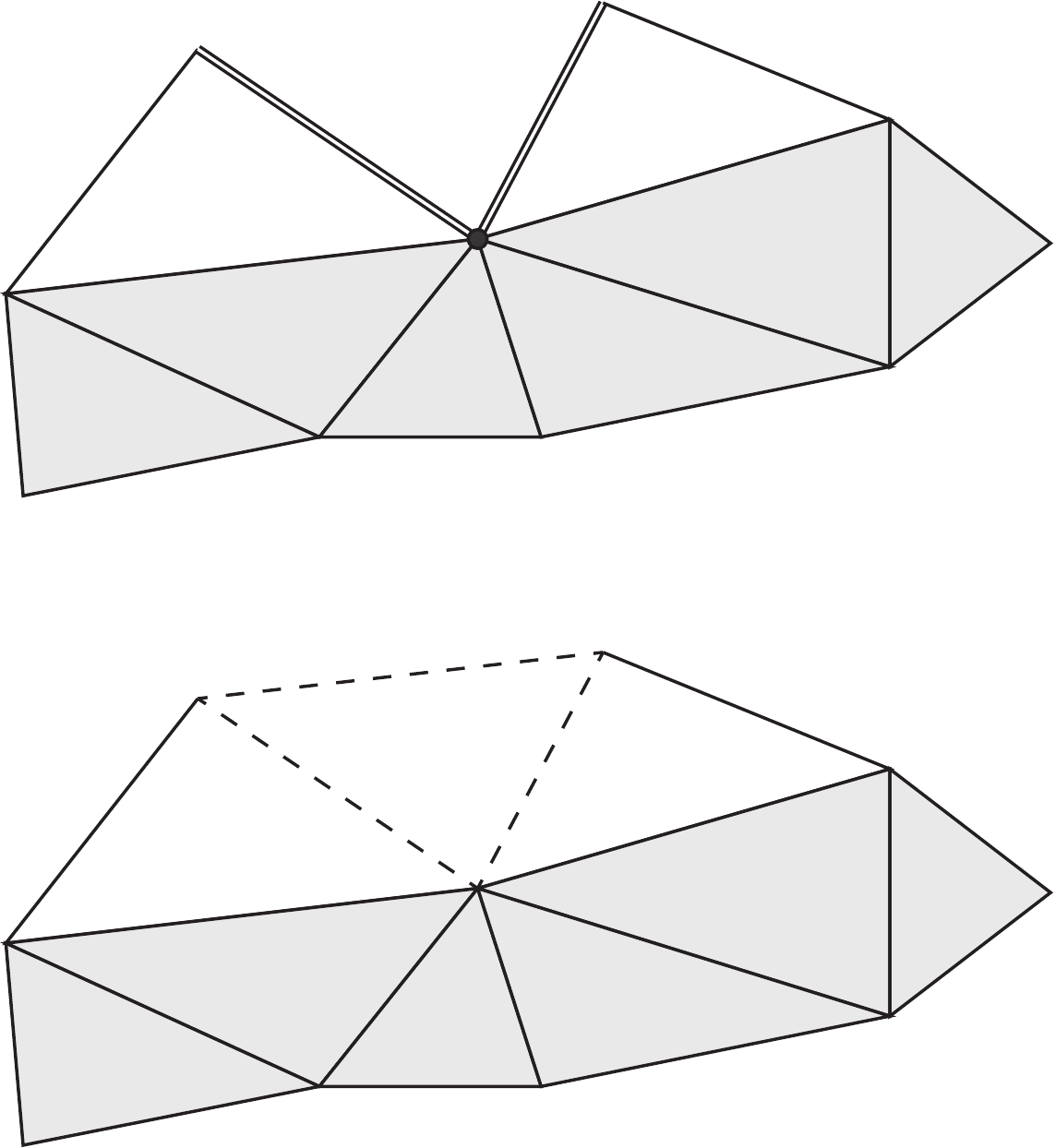}
  \caption{Gap filling. \textbf{Top}: mesh surface with the common vertex (\textbf{$\bullet$}) and free edges (\textbf{=}) indicated. \textbf{Bottom}: mesh with the new triangle indicated.}
  \label{fig:gap_filling}
\end{figure}

Gap filling is performed among the triangles used in reconstruction (\textit{$I_{R}$}). More specifically, if two reconstruction triangles share a common vertex but not a common edge (i.e. each has at least a free edge) then a new face is created, which connects the common vertex of the faces and the two corresponding adjacent free edges (see Fig.~\ref{fig:gap_filling}). The new triangle is added to the \textit{$I_{R}$} set and the process is repeated until no free edges exist.

Intuitively, the gap filling process iteratively creates new faces that complete the free edges of the existing ones until no free edges exist anymore.

% OK TODO: *** add image showing this

The gap filling algorithm is outlined in Algorithm~3. %***\ref{alg:gap_filling}.

The result of \textbf{Mesh Repair} combined with the \emph{unfolded} connected components of \textbf{Connected Component Processing} is the final output of the proposed method (See Fig.~\ref{fig:folding_step_05}).

\section{Experimental Evaluation}
\label{sec:results}

\subsection{Qualitative Results}

%In this section, the experimental evaluation of the proposed mesh folding recognition and repair method is presented.
Since no quantitative benchmark exists for the specific type of problem, a qualitative evaluation will be performed based on the quality of the method's output mesh surfaces in comparison to the input 3D models.

In order to make the results of the proposed method comparable against existing and future works on the problem of 3D mesh folding, the experiments have been performed on publicly available 3D models from well known repositories. The selection of the specific models was made based on the diversity of their shape and surface characteristics. The models used are the following:

\begin{description}
\setlength\itemsep{1em}

\item \textbf{Stanford 3D Scanning Repository}: bunny, dragon
%\item [Stanford Dragon]
%\item [Stanford Armadillo]
\item \textbf{Purdue Engineering Shape Benchmark}: m367, m826
\item \textbf{Princeton Shape Benchmark}: m545
%\item [TOSCA]
\item \textbf{McGill 3D Shape Benchmark}: b20, b234, b103, b376
\item \textbf{SHREC'13 Track: Large-Scale Partial Retrieval Using Simulated Range Images}: T169
\end{description}
%*** ΟΚ TODO add the models list
%*** ΟΚ TODO add image(s) showing this... also fix the reference in text. TODO show images for the 3D models

A mesh deformation operator (offsetting) is first applied to each 3D model, to the extent that it introduces foldings on the mesh surface. The amount of offsetting applied varies across models, since they originate from various datasets and have different resolutions.
For each model three stages are illustrated in the results figures: (a) the original 3D model, (b) the deformed 3D model (with foldings) and (c) the repaired 3D model, after its has been processed by the proposed method. In (b), the intersecting faces (\textit{I}) of the sample 3D models are also indicated. To facilitate comparative future studies, the 3 versions of all models are provided for download in the following link: ***.

It should be noted that once a 3D model has been deformed by an offsetting operator, it is not possible to revert it to its original state \cite{kestin1976second}. The purpose of the proposed method is to repair the mesh surface at the deformed state.

\begin{figure*}
  \centering
  \includegraphics[width=0.9\linewidth]{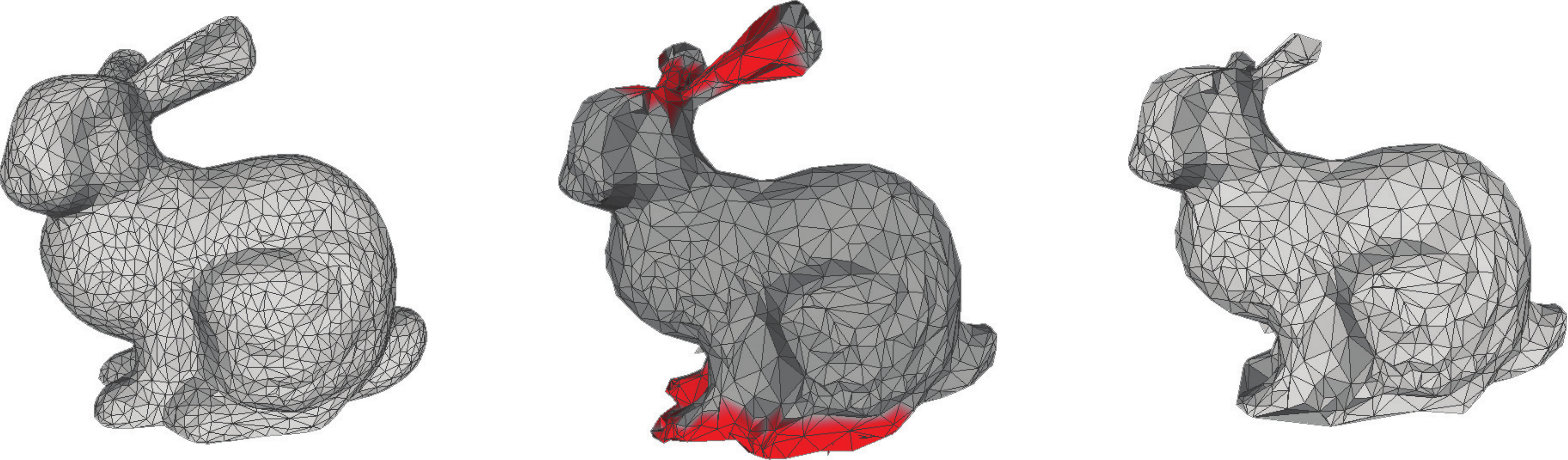}
  \includegraphics[width=0.9\linewidth]{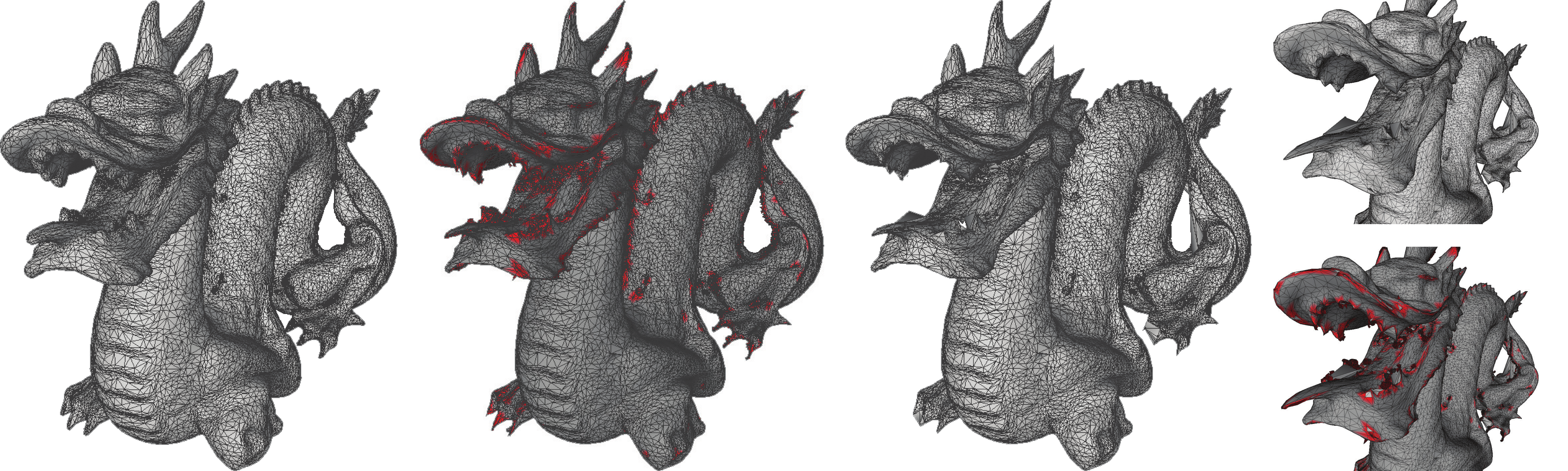}
  \caption{Illustration of the selected experimental models from the Stanford 3D Scanning Repository. \textbf{Top}: the bunny model and \textbf{bottom - left}: the dragon model. The leftmost model illustrates the original 3D mesh surface; the middle model illustrates the deformed mesh surface with red areas indicating intersecting faces; the right model illustrates the repaired mesh surface by the proposed method.  \textbf{Bottom-Right}: detail of the dragon model before and after repair. }
  \label{fig:exp_stanford}
\end{figure*}

Fig.~\ref{fig:exp_stanford} shows the results of the proposed method on the well-known bunny and dragon models of the Stanford 3D Scanning repository. The bunny model~\cite{Turk:1994:ZPM:192161.192241} exhibits intense foldings on its protruding regions (the ears) and on its base where the complexity of the surface is higher. The faces on the tips of the ears have their orientation reversed while the base faces are intersecting. The proposed method completely removed the problematic regions and reconstructed them using the retained structure. Most of the ear faces are cut off and the remaining (smaller) regions are stitched. In the base of the 3D model, the intersecting faces are removed and the reconstruction resulted in an approximation of the original structure.
The dragon model~\cite{Curless:1996:VMB:237170.237269} is of higher resolution and the foldings are less intense but spread throughout its surface. It is clear that foldings mainly occur in regions of high detail. Here, the \emph{folded} regions are removed and the holes created are reconstructed using the corresponding reconstruction faces (\textit{$I_{R}$}). Detail of the results compared with the deformed mesh surface is also given in the inset figure.

\begin{figure}
  \centering
  \includegraphics[width=0.9\linewidth]{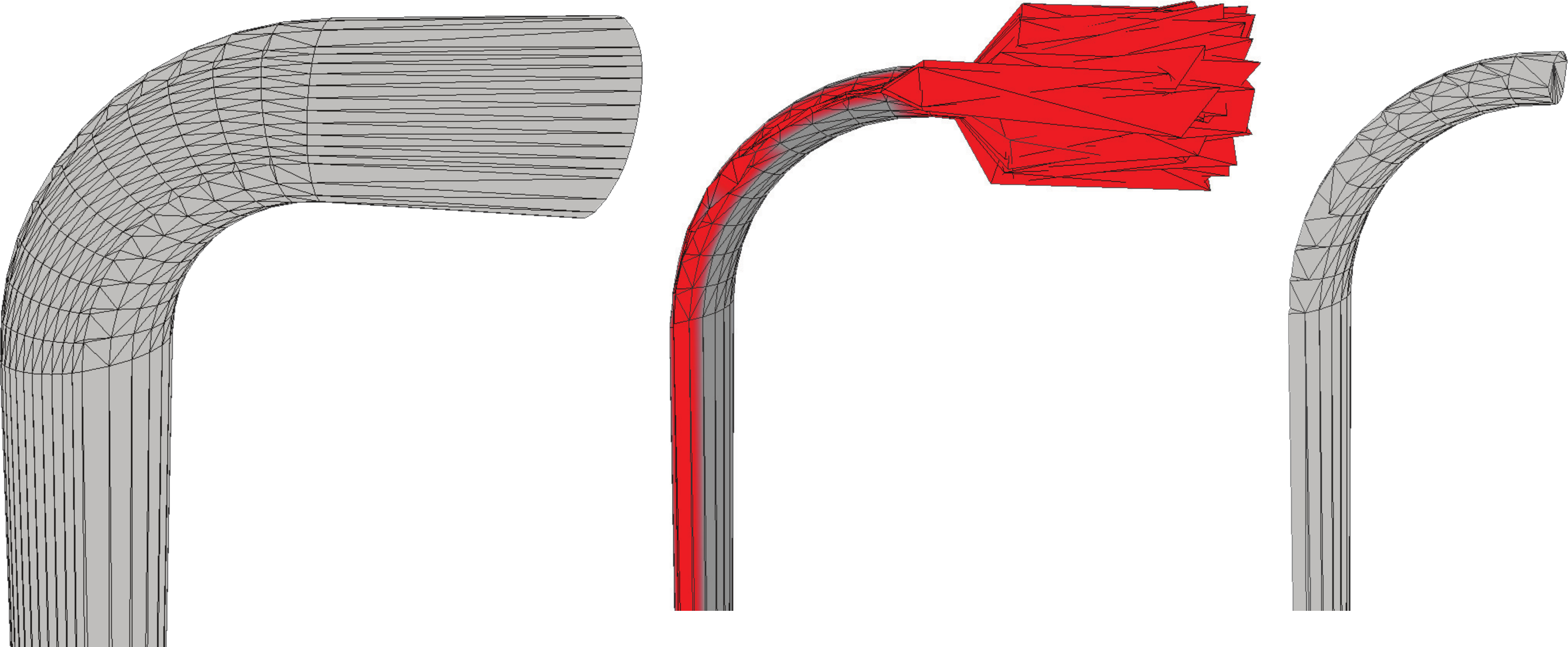}
  \includegraphics[width=0.9\linewidth]{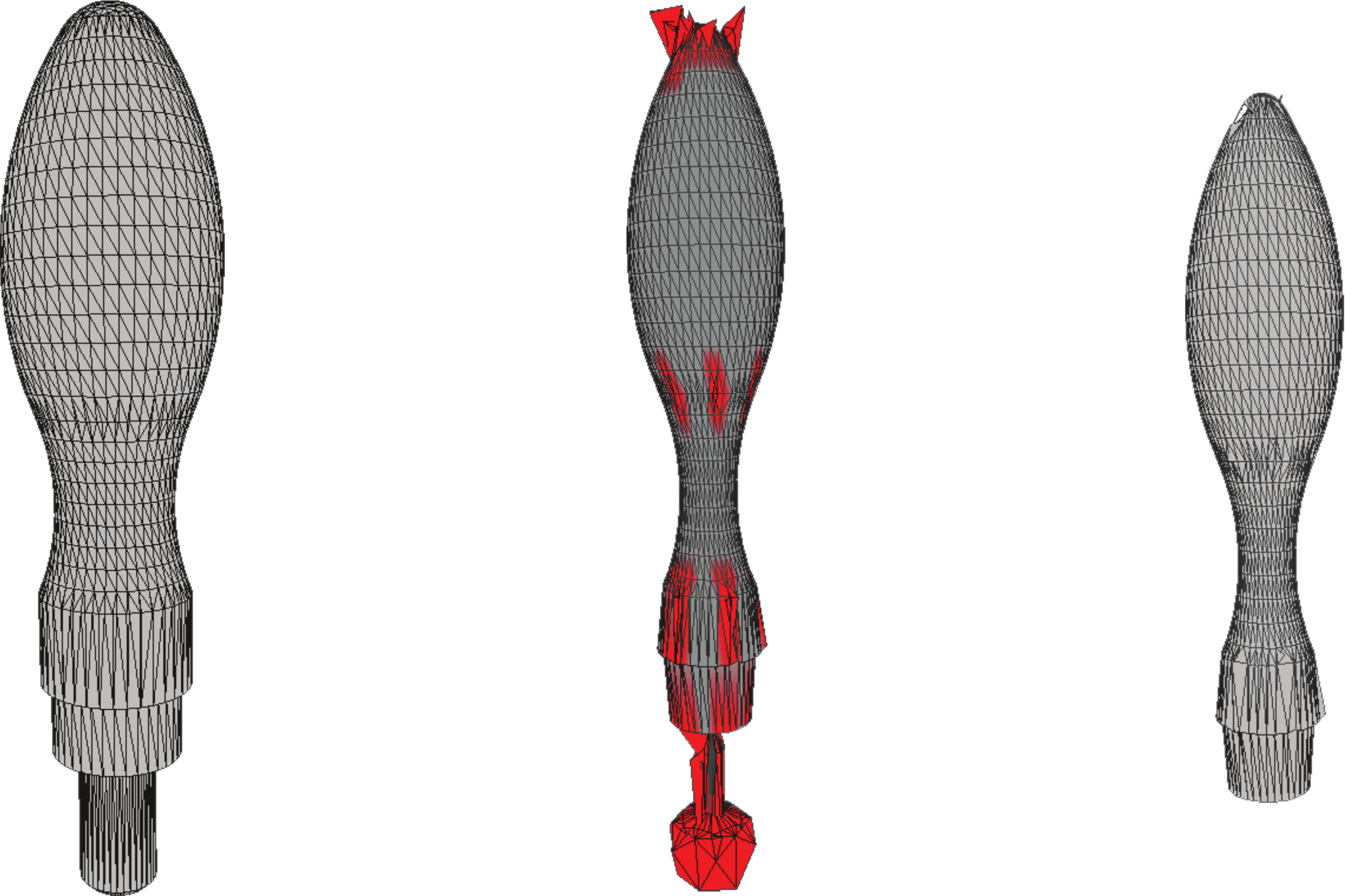}
  \caption{Illustration of the selected experimental models from the Purdue Engineering Shape Benchmark. \textbf{Top}, model m367 and \textbf{bottom}, model m826.}
  \label{fig:exp_esb}
\end{figure}

The Purdue Engineering Shape Benchmark (ESB)~\cite{jayanti2006developing} contains 3D models of CAD objects originating mainly from engineering parts. Engineering parts typically have high genus, rounding features (fillets, chamfers), presence of internal structure. They are closed watertight volumes. The application of an offsetting operator on these models can abruptly alter their shape. Here, two selected ESB models are illustrated (Fig.~\ref{fig:exp_esb}).
Model m367 of the dataset is a tubular object with rounding features. The offsetting operator has resulted in the thinning of the model's shape, creating \emph{folded} regions, especially at the endpoints. This behaviour can be explained by the difference of structure along the mesh surface. The proposed method was able to completely remove the \emph{intersecting} part of the mesh surface, resulting in a 3D model that resembles, in terms of structure, its original surface.
Model m826 contains regions of high curvature as well as abrupt edges. It is clear that the offsetting operator induced foldings in these regions. Various protruding faces have been created. The proposed method removed the faces of the \emph{foldings} and reconstructed the surface. The protruding faces that were present have been removed (observe the top of the model). The base of the 3D model, where most of the abrupt edges are located, was severely deformed, resulting in its removal.

\begin{figure}
  \centering
  \includegraphics[width=0.9\linewidth]{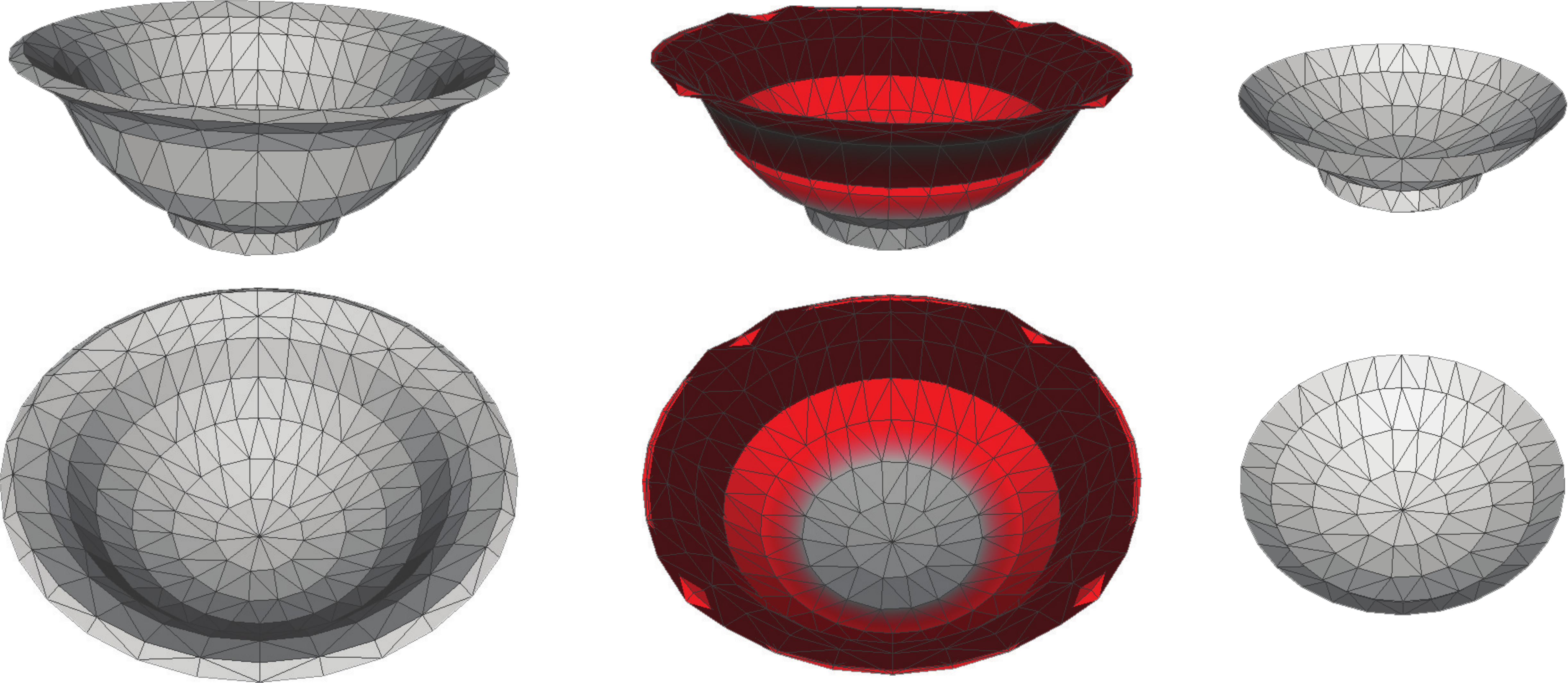}
  \caption{Illustration of the selected experimental model from the Princeton Shape Benchmark. Model m545. Front and side views are illustrated.}
  \label{fig:exp_psb}
\end{figure}

The popular Princeton Shape Benchmark (PSB)~\cite{shilane2004princeton} contains 3D models of low resolution used extensively to benchmark 3D object retrieval methodologies. To demonstrate robustness of the proposed method with respect to the resolution of the 3D models, a characteristic 3D model from PSB has been selected. The thin structure of the 3D model resulted in creating intersections and reversing the orientation of its faces. The proposed method completely removed the \emph{folded} areas and retained the \emph{unfolded} region, Fig.~\ref{fig:exp_psb}.

\begin{figure}
  \centering
  \includegraphics[width=0.9\linewidth]{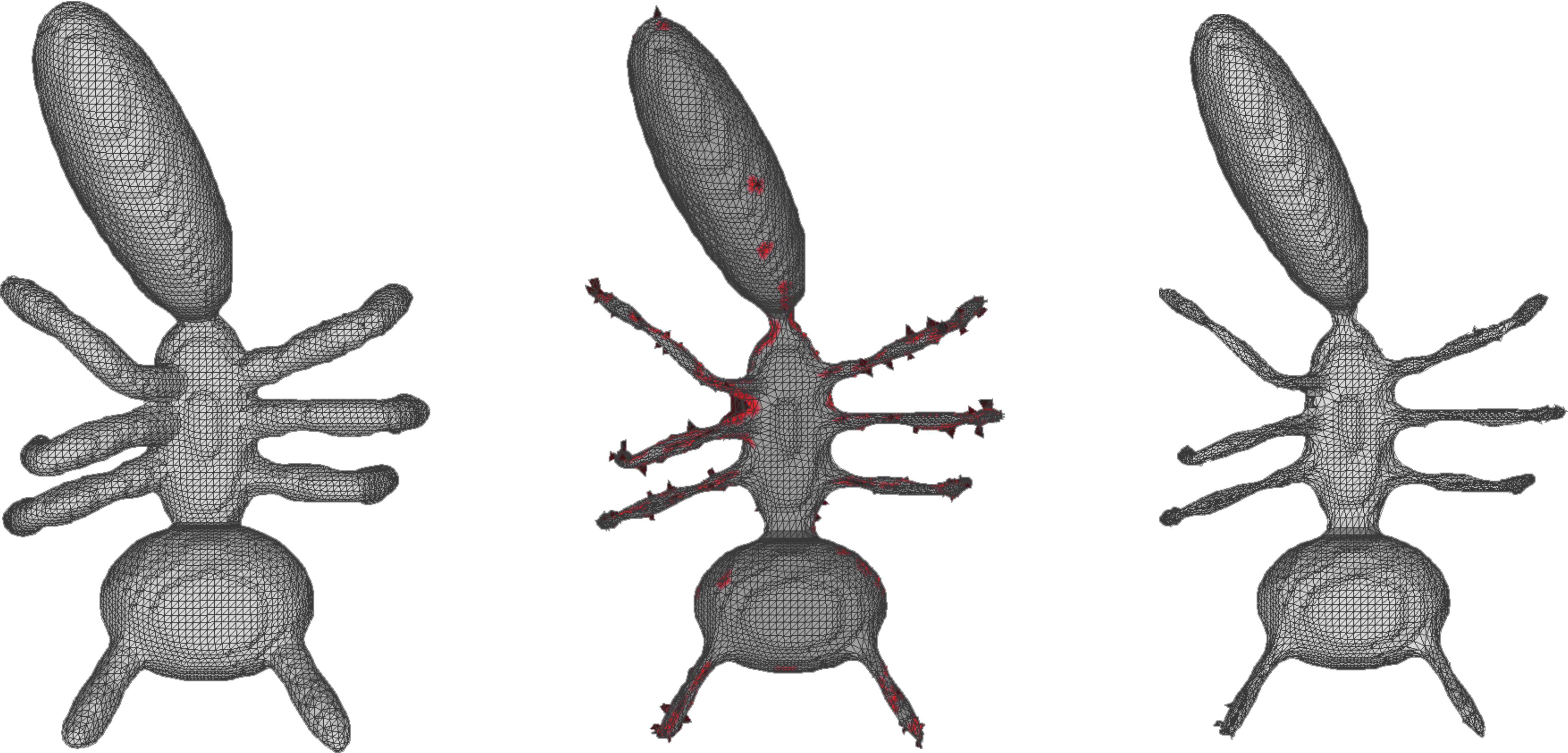}
  \includegraphics[width=0.9\linewidth]{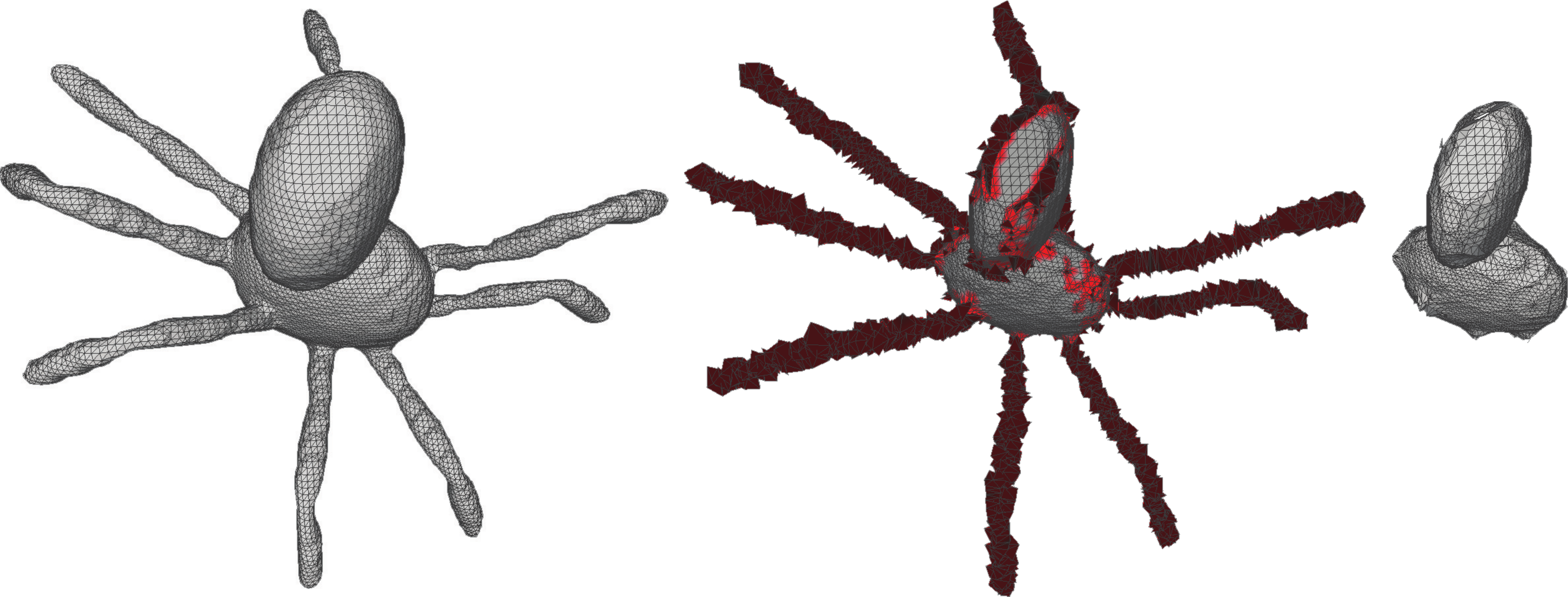}
  \includegraphics[width=0.9\linewidth]{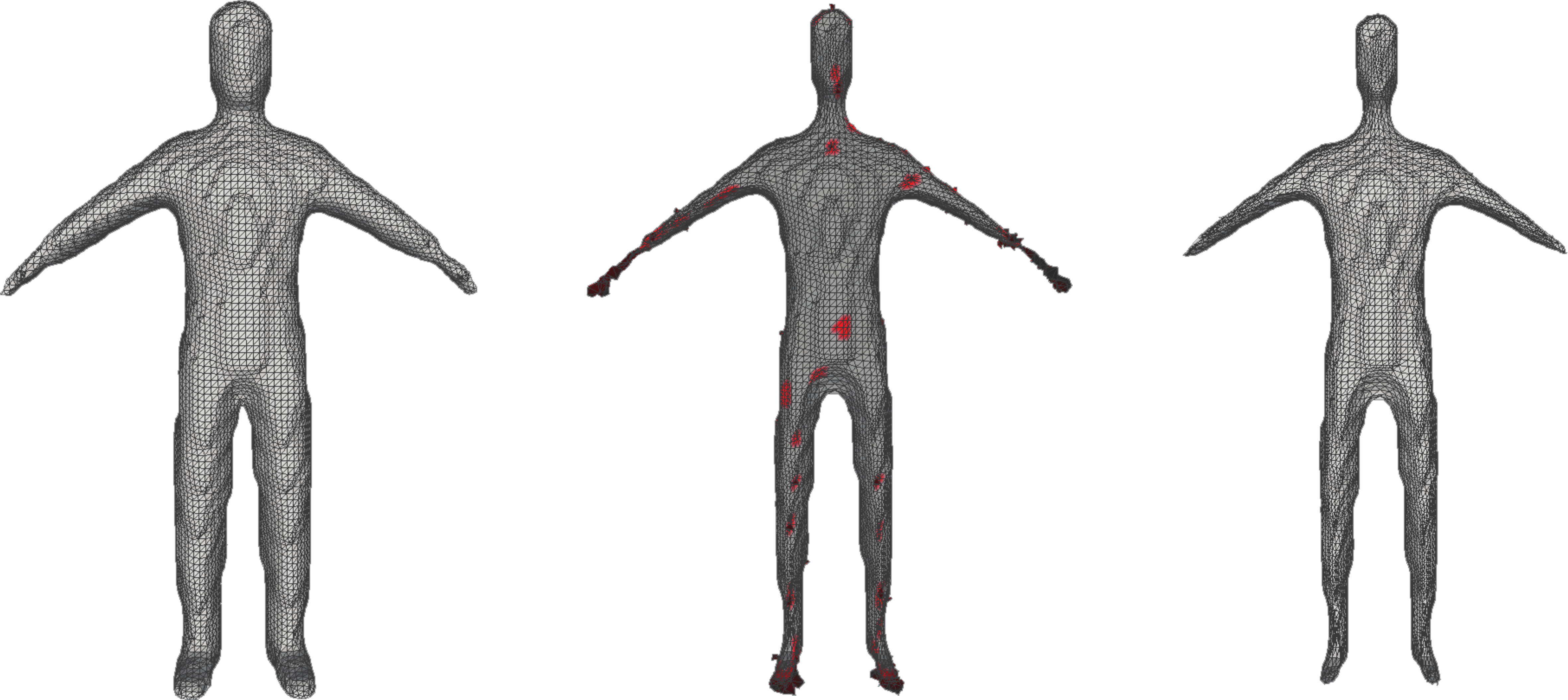}
  \includegraphics[width=0.9\linewidth]{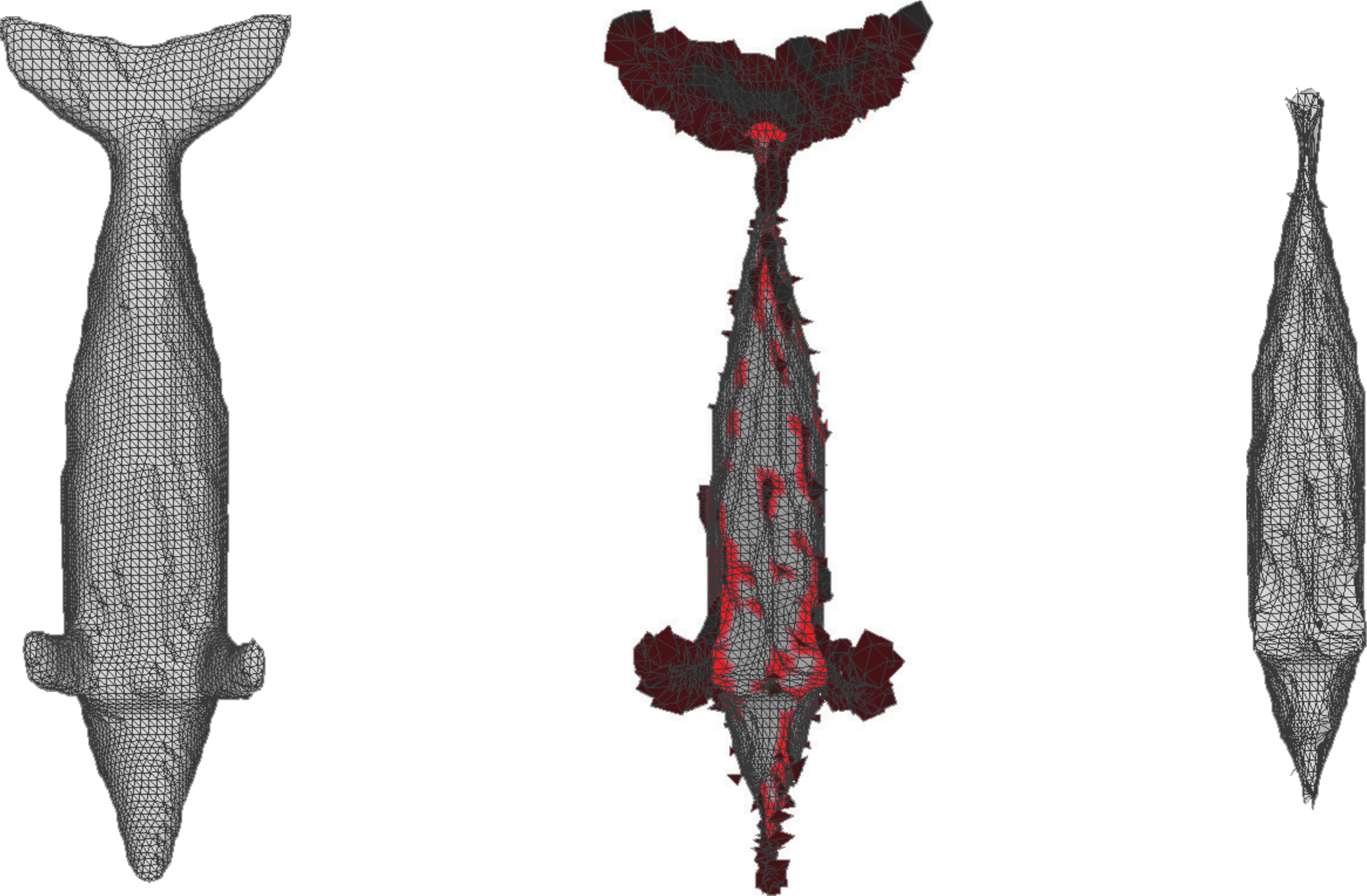}
  \caption{Illustration of the selected experimental models from the McGill 3D Shape Benchmark. From top to bottom: models b20, b234, b103 and b367.}
  \label{fig:exp_mcgill}
\end{figure}

Four 3D models have been selected from the McGill 3D Shape Benchmark~\cite{Siddiqi:2008:RAM:1390056.1390061}, demonstrating various effects of the offsetting operator on the mesh surface, as well as how these are handled by the proposed method.
Models b20 and b234 are of similar shape, having a core body and a number of articulations. On model b20, the offsetting operator resulted in several minor foldings on the mesh surface that are efficiently treated by the proposed method without any severe loss of structure. However, on model b234 the extent of the deformations was such, that the articulations were completely destroyed. The proposed method removed the problematic areas and retained the main body of the 3D model. The holes created by the removal were closed.
Application of the offsetting operator on model b103 results in a thinning of its structure that in turn severely deforms the parts of the mesh surface which were already thin (hands and tips of feet). These regions have been successfully recognized by the proposed method and removed. The resulting holes were reconstructed and closed.
The last model from the McGill dataset (b367) has undergone severe surface deformation due to the application of the offsetting operator, both on its protruding regions, as well as on its core body. The proposed method recognized the \emph{folded} regions and successfully removed them. Any gaps created by the removals (both of the protruding regions and the \emph{folded} faces of the core body) have been efficiently closed by reconstruction and filling. See Fig.~\ref{fig:exp_mcgill}.

\begin{figure}
  \centering
  \includegraphics[width=0.9\linewidth]{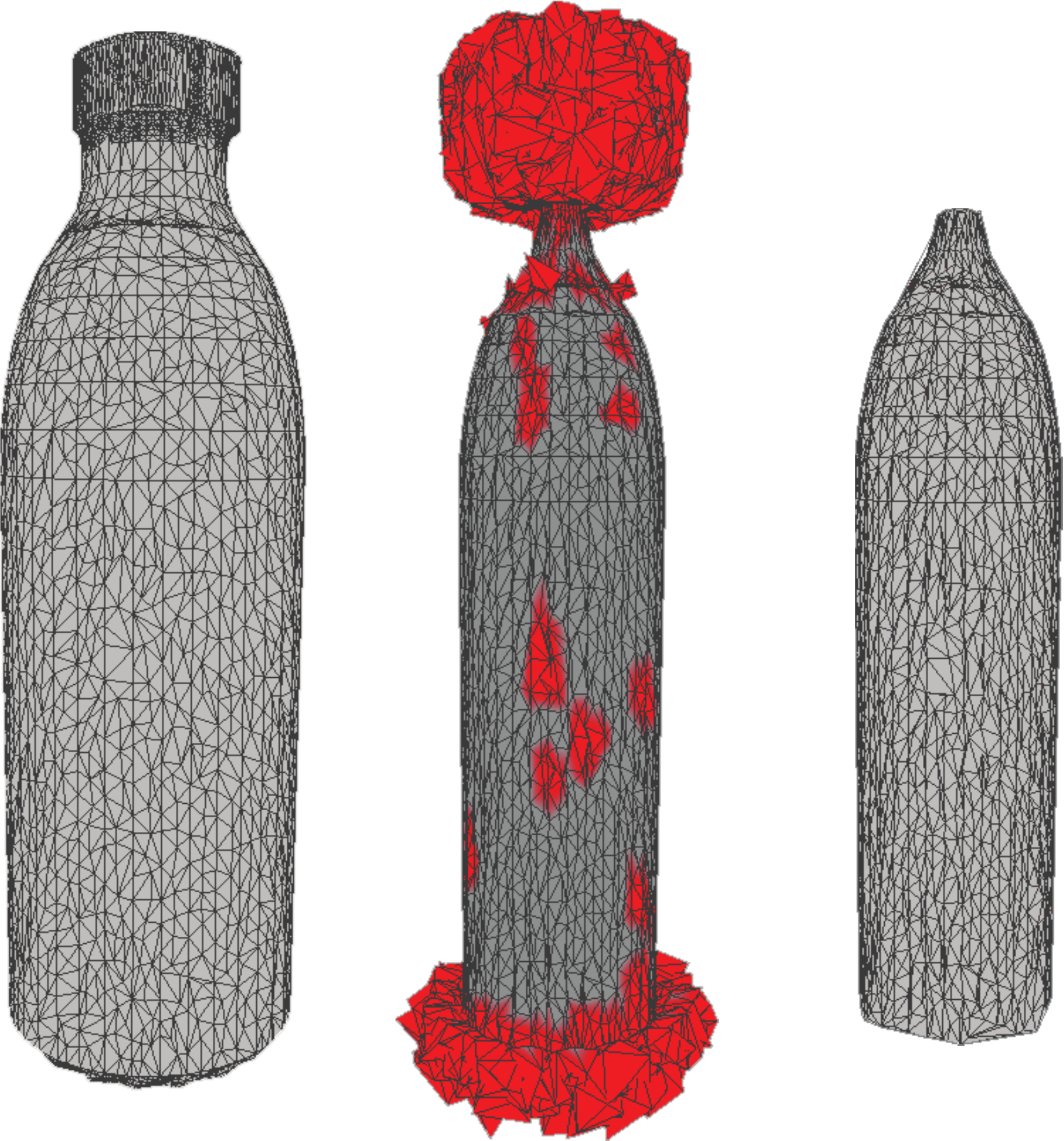}
  \caption{Illustration of the selected experimental model from the SHREC'13 Track: Large-Scale Partial Retrieval Using Simulated Range Images. Model t169.}
  \label{fig:exp_shrec}
\end{figure}

The final experimental model has been selected from the SHREC'13 Track: Large-Scale Partial Retrieval Using Simulated Range Images~\cite{sipiran2013shrec}, containing models mainly for the purposes of partial 3D object retrieval. Here, one of the complete target subset 3D models has been selected (t169). The abrupt edge regions of the 3D model exhibit the most foldings, resulting in high levels of surface destruction. The proposed method was able to remove the problematic areas completely and to reconstruct the \emph{unfolded} mesh surface. The loss of information by the removal of the \emph{folded} regions rendered impossible the accurate reconstruction of the mesh surface. However the result retains most of its original structure. See Fig.~\ref{fig:exp_shrec}.

In all the presented experiments, the mesh deformations resulted in moderate to severe alteration of the 3D models' structure. In many cases, the loss of information was of such extent that it is impossible to comprehend the shape of the object. The proposed method was able to recognize and remove the \emph{folded} regions. The gaps created by the removal process were repaired efficiently, using as structuring elements the \emph{intersecting} faces of the 3D model. In all cases, the results are robust, regardless the diversity of the 3D models in terms of structure, resolution and characteristics.

\subsection{Computational Complexity}
\label{sec:comp_comp}
%Following the qualitative evaluation of the proposed method, a brief analysis of the computational complexity with respect to the method's performance will be carried out. Due to the extent of the algorithm's structure, the complexity analysis will be performed on the key components of the pipeline.
In the following, $n$ denotes the number of vertices and $m$ denotes the number of faces of a given mesh surface.

The \textbf{mesh intersection processing} step involves the detection and removal of mesh surface intersecting faces \textit{I}. For the intersection detection task, two consecutive strategies are employed, a global intersection detection step that involves a bounding volume test and a local intersection detection step that checks only candidate triangle pairs indicated from the global intersection detection step. The global intersection detection step has a complexity of $O(m^2)$ and the local intersection detection step has a worst case cost of $O(m^2)$. %and an average run-time of $O(n^2/4)$.

The run-time of the complete mesh intersection processing step of the proposed method is similar, in terms of complexity, to the acclaimed method proposed by M\"{o}ller \cite{Moller:1997:FTI:272317.272320}, with one key difference: the cost of the operators employed.
M\"{o}ller's method is quite expensive, as it performs three subtractions, one addition, one cross product and two dot product operations, per triangle comparison.
The proposed method, stores the minimum and maximum vertex values of each dimension for every triangle. These values are then used in each axis-aligned bounding volume intersection test, resulting in six comparisons and five logical operations.
Furthermore, as it has already been defined, in near planar surfaces, M\"{o}ller's global intersection detection step results in a large number of candidate triangle pairs which must be evaluated in the local intersection detection step, thus converging to its worst run-time more frequently than the proposed method.
Once the intersection detection step is complete, the removal process can be completed in linear time O(m).

The \textbf{connected components processing} step is bounded by the connected components partitioning method %(see Algorithm~\ref{alg:non_rec_conn_comp}).
Since every triangle is tested against every other triangle for adjacency, complexity is $O(m^2)$. The algorithm is implemented in a non-recursive fashion, due to the large number of faces that a 3D model could potentially have and the restrictions imposed by stack size. In cases where recursive implementations are feasible, known algorithms of lower complexity (i.e. O(m+n)) could be employed \cite{journals/siamcomp/Tarjan72,Aho:1983:DSA:577958}.

The last step of the proposed method, \textbf{mesh repair}, is based on a (limited) number of intersecting faces. The complete mesh repair step (gap reconstruction and gap filling) has a worst case complexity of $O(m^2)$. The triangle splitting algorithm, predominately used in this step, also has a complexity of $O(m^2)$.

The actual run-time for the complete FoR$^2$M method on a typical small scale $\sim$1,000 vertex/ $\sim$2,000 face 3D model is roughly $\leq 500$~ms. For a larger 3D model ($\sim$50,000 vertices/ $\sim$100,000 faces) the run-time is approximately $\sim$150,000~ms ($\sim$2.5~min). Furthermore, to test the efficiency of the proposed triangle intersection test over M\"{o}ller's algorithm, the corresponding module was swapped for M\"{o}ller's. On the small scale 3D model, the execution time was $\sim$1,500ms, or an increase of 3 times, while on the larger 3D model the method was unable to complete, probably due to the inability of the implementation used to handle this amount of data.

All experiments were performed on an Intel (R) Core (TM) i7 @ 3.60GHz system, with 16GB of RAM. The proposed method was implemented in pure C++.

\section{Conclusion}
\label{sec:conclusions}

A novel method for the recognition and repair of foldings in 3D surface models, was presented. The method operates efficiently using a pipeline of operators on the mesh surface, requiring a simple mesh representation as input, while it does not perform any type of embedding of the input (i.e. voxelization), that could potentially alter its level of detail.
Initially, intersecting faces of the foldings are detected and removed, thus resulting in a set of connected components. The proposed intersection detection algorithm is proven to be more efficient than current approaches in terms of speed and scalability.
Next, the connected components are characterized as either \emph{folded} or \emph{unfolded}, based on the orientation of their faces. The components that have their faces \emph{inward oriented} are removed and the retained regions are stitched together by reconstructing the gaps among them. To facilitate the reconstruction process, intersecting faces of the original mesh surface are split and used as elements.
The output of the proposed method preserves the structure of the input 3D model by retaining the entire \emph{unfolded} and by using the intersecting faces of the original 3D model in the gap reconstruction process. The proposed method is efficient in terms of speed both as a whole and in terms of its individual processes.
Target of the proposed method is to facilitate mesh degradation procedures (i.e. erosion) so that the output 3D mesh maintains the corresponding structure. In this fashion mesh removal and reconstruction/filling procedures were employed.

% if have a single appendix:
%\appendix[Proof of the Zonklar Equations]
% or
%\appendix  % for no appendix heading
% do not use \section anymore after \appendix, only \section*
% is possibly needed

% use appendices with more than one appendix
% then use \section to start each appendix
% you must declare a \section before using any
% \subsection or using \label (\appendices by itself
% starts a section numbered zero.)
%

\appendix[Pseudocode References]
\label{sec:appendix}

\begin{algorithm*}[!ht]
\caption{Triangular Faces Intersection Detection and Splitting}
\label{alg:tri_split}
\begin{algorithmic}[1]
\Require $F$                                                                            \Comment{input face triangles set}
\Ensure $F_{split}$                                                                     \Comment{output split face triangles set}
\State IsectList = [];                                                                  \Comment{intersection list for each triangle}

\For {$i = 0$ to length($F$)}                                                           \Comment{for every triangle in set $F$}
    \For {$j = 0$ to length($F$)}                                                       \Comment{for every triangle in set $F$}
        \If {Intersects($F[i]$, $F[j]$)}                                                \Comment{test if the triangle $j$ intersects triangle $i$}
            \State IsectList$[i]$.push$[j]$;                                            \Comment{push back triangle $j$ in triangle $i$ intersection list}
        \EndIf
    \EndFor
\EndFor

\For {$i = 0$ to length(IsectList)}                                                     \Comment{for every \emph{intersected} triangle}
    \State SplitTriangleList = $F[i]$;                                                  \Comment{Initialise split triangle list to this triangle}
    \For {$j = 0$ to length(IsectList[$i$])}                                            \Comment{for every \emph{intersector} triangle}
        \For {$k = 0$ to length(SplitTriangleList)}                                     \Comment{for every \emph{intersected} base triangle}
            \If {TriangleSplitsTriangle(IsectList$[i][j]$,SplitTriangleList$[k]$,SplitResultFaces)}        \Comment{if \emph{intersector} splits \emph{intersected} then...}
                \State SplitTriangleList.pop($k$);                                      \Comment{remove current triangle from base and...}
                \State SplitTriangleList.push(SplitResultFaces);                        \Comment{push 3 new triangles into base}
                \State $k = k - 1$;                                                     \Comment{repeat procedure for the remaining of the (updated) base}
            \EndIf
        \EndFor
    \EndFor
    \State $F_{split}$.push(SplitTriangleList);                                         \Comment{output is the split list of \emph{intersected} triangles}
\EndFor
\end{algorithmic}
\end{algorithm*}

\begin{algorithm*}[!ht]
\caption{Gap Reconstruction}
\label{alg:gap_reconstruction}
\begin{algorithmic}[1]
\Require $F_{split}$                                                                    \Comment{input split face triangles set; triangles are arranged in triplets, so that the first triangle of each triplet is the triangle that lies single on one side of the intersection line. The remaining two triangles lie on the other side of the intersection line.}
\Require ConnComp                                                                       \Comment{input connected components structure of the 3D model}
\Ensure $F_{recon}$                                                                     \Comment{list of output faces used to reconstruct the gaps between the connected components}

\For {$i = 0$ to length($F_{split}$), step 3}                                           \Comment{for every triangle in set $F_{split}$}
    \If {IsAdjacent($F_{split}[i]$, ConnComp)}                                          \Comment{if first triangle is adjacent to any of the connected components}
        \State $F_{recon}$.push($F_{split}[i]$);                                        \Comment{use it for gap reconstruction}
    \ElsIf {IsAdjacent($F_{split}[i + 1]$, ConnComp) or IsAdjacent($F_{split}[i + 2]$, ConnComp)}
    \\                                                                                  \Comment{if second or third triangle is adjacent to any of the connected components}
        \State $F_{recon}$.push($F_{split}[i + 1]$);                                    \Comment{use both of them for gap reconstruction}
        \State $F_{recon}$.push($F_{split}[i + 2]$);
    \EndIf
\EndFor
\end{algorithmic}
\end{algorithm*}
\ifCLASSOPTIONcaptionsoff
  \newpage
\fi

\newpage

% trigger a \newpage just before the given reference
% number - used to balance the columns on the last page
% adjust value as needed - may need to be readjusted if
% the document is modified later
%\IEEEtriggeratref{8}
% The "triggered" command can be changed if desired:
%\IEEEtriggercmd{\enlargethispage{-5in}}

% references section

% can use a bibliography generated by BibTeX as a .bbl file
% BibTeX documentation can be easily obtained at:
% http://mirror.ctan.org/biblio/bibtex/contrib/doc/
% The IEEEtran BibTeX style support page is at:
% http://www.michaelshell.org/tex/ieeetran/bibtex/
\bibliographystyle{IEEEtran}
% argument is your BibTeX string definitions and bibliography database(s)
\bibliography{./bibliography/cites,./bibliography/SelfFoldingPapersTP}
\end{document}